\documentclass[a4paper,10pt,twoside=false]{scrartcl}

\let\arxiv\undefined \def\arxiv{}  

\ifdefined\arxiv
\else
\fi

\usepackage{graphicx}
\usepackage[export]{adjustbox}
\graphicspath{{./tikz}{./fig}}
\usepackage[centertags]{amsmath}
\usepackage{amssymb}
\usepackage[english]{babel}
\usepackage{siunitx}

\usepackage[utf8]{inputenc}
\usepackage{csquotes}
\usepackage[T1]{fontenc}
\usepackage[kerning=true,protrusion=true,expansion=true]{microtype}
\usepackage[left=4.4cm,right=1.6cm,top=2.4cm,bottom=2.4cm]{geometry}
\usepackage{abstract}
\setcapwidth[c]{\textwidth}
\setcapindent{0pt}

\usepackage{bm}
\usepackage[mleftright]{diffcoeff}
\usepackage{mathtools}
\usepackage[mathscr]{euscript}

\usepackage{hyperref}
\usepackage[dvipsnames]{xcolor}
\usepackage{tikz}
\usetikzlibrary{calc}
\usepackage{tcolorbox}
\tcbuselibrary{breakable}
\definecolor{darkblue}{rgb}{0,0,.6}
\hypersetup{colorlinks=true, linkcolor=darkblue, citecolor=darkblue}
\newtcolorbox{arbox}[2][]{
  breakable,
  boxrule  = 0.5pt,
  colframe = gray!20,
  colback  = gray!10,
  coltitle = gray!20!black,
  colbacktitle = gray!20,
  fonttitle = \sffamily,
  title    = #2,
  #1,
}

\usepackage[capitalize]{cleveref}
\usepackage{cite}

\DeclareMathOperator{\Var}{Var}
\DeclareMathOperator{\diag}{diag}

\tcbset{colback=orange!5!white, colframe=orange!75!black, boxrule=0.5pt}

\newcommand*{\nlv}{\ensuremath{{\bm\sigma}}}
\newcommand*{\mvec}[1]{{\ensuremath{\mathbf #1}}}
\newcommand*{\gains}{\ensuremath{g_i}}

\newcommand*{\cG}{\ensuremath{\mathsf{G}}}
\newcommand*{\heating}{\ensuremath{\Gamma\s{m}}}
\newcommand*{\damping}{\ensuremath{\zeta}}
\usepackage{bbold}

\newcommand\sbullet[1][.5]{\mathbin{\vcenter{\hbox{\scalebox{#1}{$\bullet$}}}}}

\newcommand*{\acomm}[2]{\ensuremath{ #1 #2 + #2 #1 }}
\newcommand*{\dyad}[2]{\ensuremath{ \mleft\lvert #1 \middle\rangle\middle\langle #2 \mright\rvert }}
\newcommand*{\dd}{\ensuremath{\mathrm{d}}}

\DeclarePairedDelimiter{\ket}{\lvert}{\rangle}\DeclarePairedDelimiterX\innerp[2]{\langle}{\rangle}{#1\delimsize\vert\mathopen{}#2}\DeclarePairedDelimiterX\braket[2]{\langle}{\rangle}{#1\delimsize\vert\mathopen{}#2}\DeclarePairedDelimiterX\braketOP[3]{\langle}{\rangle}{#1\,\delimsize\vert\,\mathopen{}#2\,\delimsize\vert\,\mathopen{}#3}\DeclarePairedDelimiterX\ketbra[2]{\lvert}{\rvert}{#1\delimsize\rangle\!\delimsize\langle#2}\DeclarePairedDelimiterX\outerp[2]{\lvert}{\rvert}{#1\delimsize\rangle\!\delimsize\langle#2}\DeclarePairedDelimiterX\projector[1]{\lvert}{\rvert}{#1\delimsize\rangle\!\delimsize\langle#1}

\DeclarePairedDelimiterX\comm[2]{[}{]}{#1,#2}\DeclarePairedDelimiter{\ev}{\langle}{\rangle}

\colorlet{acolor}{orange!90!black}

\newcommand*{\scrpt}[1]{\mathsf{#1}}
\newcommand*{\s}[1]{\ensuremath{_\scrpt{#1}}}
\newcommand*{\up}[1]{\ensuremath{^\scrpt{#1}}}
\newcommand{\ii}{\mathrm{i}}
\newcommand{\ee}{\mathrm{e}}

\newcommand*{\opr}[1]{\ensuremath{\hat{ #1 }}}
\newcommand*{\UU}{\ensuremath{\opr{\mathscr{U}}}}

\newcommand*{\xs}{\opr{q}}
\newcommand*{\xt}{\opr{X}}
\newcommand*{\ps}{\opr{p}}
\newcommand*{\pt}{\opr{Y}}

\def\BE{\begin{equation}}
\def\EE{\end{equation}}
\def\BY{\begin{align}}
\def\EY{\end{align}}

\newcommand{\thetitle}{Broadcasting quantum nonlinearity in hybrid systems}

\begin{document}
\setkomafont{author}{\sffamily}

\title{\thetitle}

\newcommand*{\UPOL}{Department of Optics, Palack{\'y} University, 17.~Listopadu~12, 771~46~Olomouc, Czech~Republic}

\author{
    Alisa D. Manukhova
    \footnote{Corresponding author: \href{alisamanukhova@gmail.com}{alisamanukhova@gmail.com,alisadmitrievna.manukhova@upol.cz}}{ }$^{1}$
    Andrey A. {Rakhubovsky}
    \footnote{\href{mailto:rakhubovsky@optics.upol.cz}{rakhubovsky@optics.upol.cz}}{ }$^{1}$
    \and
    Radim {Filip}\footnote{\href{mailto:filip@optics.upol.cz}{filip@optics.upol.cz}}{ }$^{1}$
\\
$^1$\emph{\UPOL}}

\date{}

\renewcommand{\abstractname}{}

\maketitle

\begin{abstract}
  Linear oscillators contribute to most branches of contemporary quantum science.
  They have already successfully served as quantum sensors and memories, found applications in quantum communication, and hold promise for cluster-state-based quantum computing.
  To master universal quantum processing with linear oscillators, an unconditional nonlinear operation is required.
  We propose such an operation using light-mediated interaction with another system that possesses a nonlinearity equivalent to more than a quadratic potential.
  Such a potential grants access to a nonlinear operation that can be broadcast to the target linear system.
  The nonlinear character of the operation can be verified by observing adequate negative values of the target system's Wigner function and the squeezing of the variance of a certain nonlinear combination of the quadratures below the thresholds attainable by Gaussian states.
  We explicitly evaluate an optically levitated mechanical oscillator as a flexible source of nonlinearity for a proof-of-principle demonstration of the nonlinearity broadcasting to linear systems, for example, mechanical oscillators or macroscopic atomic spin ensembles.
\end{abstract}

\section{Introduction} 

Advanced quantum applications with linear bosonic oscillators require nonlinear quantum gates to reach the full advantage of nonlinear processing required for the universality~\cite{lloyd_quantum_1999} and, ultimately, thresholds for fault-tolerant quantum computation~\cite{braunstein_quantum_2005}.
On the one side, such challenging tasks require the oscillators to preserve linearity in order to keep essential features of continuous-variable quantum information; simultaneously, they also require specific nonlinear phase gates~\cite{lloyd_quantum_1999} appropriate for gate synthesis towards complex circuits and applications~\cite{kendon_quantum_2010,georgescu_quantum_2014}.
Such unique nonlinearities are challenging to realize at the quantum level in the advanced linear-oscillator platforms like, for example, light pulses, linear optomechanics, and optically controlled spin ensembles suitable for important storing of the bosonic states~\cite{cerf_quantum_2007,polzik_quantum_2008}.
The need to reach the nonlinearities opens space for hybrid quantum devices~\cite{treutlein_hybrid_2014,clerk_hybrid_2020,chu_perspective_2020} that can, in principle, broadcast nonlinearity from a nonlinear oscillator to the linear one.
The hybrid continuous-variable quantum connection has been recently stimulated by two parallel experiments demonstrating quantum correlations between atomic ensembles and mechanical oscillators using an optical mediation~\cite{moller_quantum_2017,karg_remote_2019,karg_lightmediated_2020,thomas_entanglement_2020}.

Quantum optomechanics~\cite{aspelmeyer_cavity_2014} studies parametric interactions of radiation with mechanical motion.
At the quantum level, linear optomechanical devices have demonstrated impressive control over quantum states of radiation and mechanics including ground-state cooling of mechanics~\cite{teufel_sideband_2011,chan_laser_2011,delic_cooling_2020,piotrowski_simultaneous_2023}, generation of squeezed states of mechanical oscillator~\cite{wollman_quantum_2015,pirkkalainen_squeezing_2015} and field~\cite{safavi-naeini_squeezed_2013,ockeloen-korppi_noiseless_2017,militaru_ponderomotive_2022}, optomechanical~\cite{palomaki_entangling_2013,riedinger_nonclassical_2016} and mechanical~\cite{riedinger_remote_2018,ockeloen-korppi_stabilized_2018} entanglement.
Mechanical oscillators naturally couple to forces of different nature which explains their wide applicability in quantum sensing~\cite{ranjit_zeptonewton_2016,dominguez-medina_neutral_2018,mason_continuous_2019}.
This has practical use and is important for studies of fundamental science~\cite{carney_mechanical_2021,carney_searches_2023,heinze_darkgeo_2024,baker_optomechanical_2024}.
Another application of optomechanics is in transduction of quantum information between light and microwaves~\cite{bagci_optical_2014,higginbotham_harnessing_2018}, which enables long-distance quantum communication between superconducting devices, and optical detection of microwave signals.

In recent years, there has been an effort toward hybridization of optomechanical systems with systems of different nature~\cite{treutlein_hybrid_2014,clerk_hybrid_2020,chu_perspective_2020}.
Such a hybridization can be beneficial for both involved sides.
Typically, the mechanical system can enjoy increased quantum control, in exchange providing the capabilities to filter, store, and transduce quantum signals.
One particularly interesting direction is coupling optomechanical systems to atoms~\cite{moller_quantum_2017,karg_remote_2019,karg_lightmediated_2020,thomas_entanglement_2020}.

Optomechanics can be helpful for the implementation of nonlinear operations for linear mechanical or atomic ensembles.
Recently, levitated nanoparticles~\cite{gonzalez-ballestero_levitodynamics_2021} have attained a flexible access to potentials of mechanical motion that can be engineered to be nonlinear beyond the quadratic potential of harmonic oscillator~\cite{siler_diffusing_2018,gutierrezlatorre_superconducting_2023}.
Using such a potential, it is possible to adjust nonlinear transformations of quantum states~\cite{rakhubovsky_stroboscopic_2021,neumeier_fast_2024,roda-llordes_macroscopic_2024,casulleras_optimization_2024} in a versatile manner.
There is, as well, an ongoing effort, both experimental and theoretical, to achieve and harness the nonlinearity of mechanical oscillations in bulk systems~\cite{peano_nonlinear_2006,leijssen_nonlinear_2017,guerra_electrostatically_2008,jacobs_engineering_2009,matheny_nonlinear_2013,rips_nonlinear_2014,ochs_amplification_2021,ochs_frequency_2022}, also see Ref.~\cite{rosiek_quadrature_2023} and references therein.

An outstanding degree of quantum control puts macroscopic atomic ensembles forward among the best candidates for quantum science~\cite{chu_cold_2002}.
Atomic ensembles have been instrumental in fundamental research, including quantum metrology and sensing~\cite{kitching_atomic_2011}.
Prominent room-temperature control, linearity, and long coherence times make atoms especially suitable for quantum information~\cite{monroe_quantum_2002,cerf_quantum_2007,lukin_colloquium_2003,kimble_quantum_2008,sangouard_quantum_2011}.
Atomic ensembles were among the first to demonstrate coherent storage and retrieval of quantum information~\cite{vanderwal_atomic_2003,julsgaard_experimental_2004,laurat_efficient_2006,simon_interfacing_2007,honda_storage_2008,appel_quantum_2008,choi_mapping_2008,yang_efficient_2016,saunders_cavityenhanced_2016,wang_efficient_2019}.
However, universal quantum information processing over continuous variables of atoms requires specific nonlinear phase gates which the linear atomic ensembles~\cite{cerf_quantum_2007} do not have~\cite{divincenzo_physical_2000,braunstein_quantum_2005}.
Use of low-temperature Bose-Einstein condensates~\cite{anderson_observation_1995,davis_boseeinstein_1995,pethick_bose_2008} beyond linearization is not compatible with the continuous-variable quantum computation model based on linear oscillators~\cite{braunstein_quantum_2005}.

In this manuscript, we propose a hybrid strategy to broadcast a quantum nonlinear transformation on an otherwise linear system.
As an example of a linear system, we consider an atomic system complemented by the nonlinear phase transformation intrinsically available to a mechanical oscillator with a nonlinearity in the form of a potential that is a higher-than-quadratic function of its position.
However, the method is applicable to nonlinear phase gates on any platform connected to light or microwaves~\cite{eriksson_universal_2024}.
The nonlinearity is broadcast from the mechanics via a sequence of pulsed optically-mediated quantum nondemolition (QND) gates~\cite{manukhova_pulsed_2020}.
These QND gates are typically available in the interaction of light with mechanics~\cite{shomroni_optical_2019,liu_quantum_2022} and with atomic ensembles~\cite{julsgaard_experimental_2001}.
The chosen sequence of operations requires only linear interactions with the atomic cloud without the need for nonlinear feedforward~\cite{marek_general_2018}.
At the end of the protocol, the quantum state of the atoms inherits a nonlinear phase gate with an enhanced nonlinearity.
For simulations, we assume the mechanical mode to have access to a cubic nonlinearity as was proposed for levitated optomechanics in~\cite{siler_diffusing_2018}.
Multiple other theoretical proposals stimulate the development of experimental mechanical nonlinearities in such systems at the quantum level~\cite{rakhubovsky_stroboscopic_2021,neumeier_fast_2024,roda-llordes_numerical_2024,roda-llordes_macroscopic_2024,casulleras_optimization_2024}.
The use of cubic nonlinearity is stimulated by its pivotal role of the first monomial nonlinearity capable of granting access to universal quantum information processing with continuous variables~\cite{braunstein_quantum_2005,budinger_alloptical_2024}.
We verify the nonlinearity by observing negative Wigner function phase space interference and comparing the variance of specially devised nonlinear quadrature of atoms against the thresholds attainable by convex mixtures of classical states or Gaussian states~\cite{moore_hierarchy_2022}.
Our methodology and this example navigate further efforts in bosonic hybrid systems to broadcast essential nonlinearity to linearized systems.

\section{Results} 

\begin{figure*}[htb]\centering
  \includegraphics[width=\linewidth]{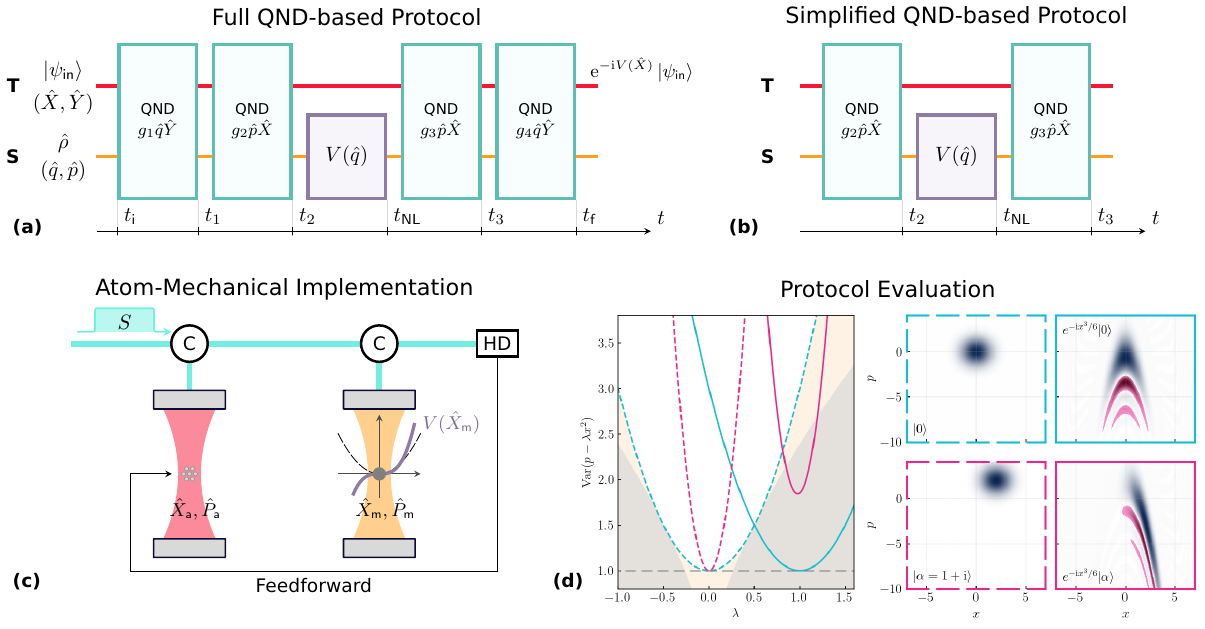}
  \caption{(a) Circuit diagram of the proposed protocol to broadcast nonlinearity from the source system (S, orange line) to the target system (T, red line).
    The protocol consists of five Hamiltonian transformations: four linear QND interactions between the systems (cyan) and a local nonlinear transformation of the source system (violet).
    The Hamiltonians corresponding to each step are written inside color boxes.
    (b) Circuit diagram of a simplified protocol with only two QND interactions.
    (c) A possible concrete implementation with a linear atomic cloud as the target system and a levitated nanoparticle as the nonlinearity source.
    Squeezed light pulse ($S$) mediates a linear QND interaction between the matter systems.
    Here, the nonlinearity is illustrated as a cubic function of the displacement of the nanoparticle.
    C --- circulators, HD --- homodyne detector.
    (d) Main figures of merit used to evaluate the broadcasting of the nonlinearity.
    Nonlinear variance~$\nlv(\lambda)$ (NLV, see~\cref{eq:nlv_definition}) evaluated for the ground state $\ket{0}$, a coherent state $\ket{ \alpha = 1 + \ii}$, and for these two states after a cubic phase gate $\exp[ - \ii \hat x^3 / 6]$.
    Filled regions of the NLV graph indicate areas reachable only by non-classical (yellow) states and quantum non-Gaussian (gray) states.
    The panels on the right show the contour plots of the Wigner functions of these states.
  }
  \label{fig:fig0-pdf}
\end{figure*}

We consider two single-mode systems, in general, a nonlinear \emph{source} one and a linear \emph{target} one.
The source system has access to a nonlinear phase gate beyond linear transformations (that are provided by the Hamiltonians at most quadratic in bosonic operators).
The target system, on the contrary, is linear, that is, its Hamiltonian is not more than quadratic.
We assume that the nonlinear phase gate can be accessed on demand and can be implemented approximately unitarily.
Furthermore, we consider that the source and target can be coupled (directly or using a mediator) via a linear quantum non-demolition (QND) interaction in a pulsed manner.
Recently, such a QND coupling has been realized between atomic ensembles and mechanical membranes~\cite{thomas_entanglement_2020,karg_lightmediated_2020}.
Below, we show how these different types of evolution can be arranged to implement an effective nonlinear transformation of the target system (which, on its own, is otherwise limited to only linear transformations).
An alternative strategy, involving a state-swap/beamsplitter-like interaction~\cite{hammerer_quantum_2010,aspelmeyer_cavity_2014} is possible, though its proper analysis is beyond the scope of the current work.
The theory is presented here in a general manner in order to be applicable to arbitrary platforms.
The numerical results of this section are obtained assuming concrete parameters inspired by and closely related to the recent experiments with atomic ensembles~\cite{karg_lightmediated_2020,thomas_entanglement_2020} and levitated nanoparticles~\cite{delic_cooling_2020}.

\subsection{Model and the quantum-non-demolition-based protocols} \label{sec:nonlinearity_distribution_to_a_linear_system_4QND}

The source and the target systems are described by the quadrature operators in the phase space, $(\xs,\ps)$ and $(\xt,\pt)$, respectively.
The quadratures of each system obey the canonical commutation relations $\comm{ \xs }{ \ps } = \comm{ \xt}{ \pt } = 2 \ii$, which corresponds to setting $\hbar = 2$.
We assume that two types of operations can be implemented on the source and target modes of these systems.
First, linear QND-type gates, described by the unitary operators
\begin{equation}
  \UU_{qy } (g) = \ee^{ - \ii g \xs \pt / 2 },
  \quad
  \text{ or }
  \quad
  \UU_{px } (g) = \ee^{ - \ii g \ps \xt / 2 }.
\end{equation}
Here, $g$ is a controllable gain of the Gaussian interaction.
In setups similar to experiments~\cite{thomas_entanglement_2020,karg_lightmediated_2020}, the gain can typically be relatively easily manipulated by the setup parameters such as drive strength, etc.

Second, the source system has a nonlinear transformation
\begin{equation}
  \label{eq:unonlinear_def}
  \UU\s{NL} (\xs, \gamma) = \ee^{ - \ii \gamma V (\xs ) / 2 },
\end{equation}
where $V$ is a nonlinear differentiable function of its argument.
For instance, for an object levitated in a cubic potential (e.g., see~\cite{siler_thermally_2017}), $\gamma V(x) \approx \alpha \tau x^3 / 3$, where $\gamma = \alpha \tau$;
$\alpha$ denotes the stiffness of the potential and $\tau$ duration of the evolution in the potential.
For simplicity, we parametrize such an evolution by a single number, $\gamma$.
The equality is approximate because to reach the gate, we assumed a short evolution time and disregarded the influence of the kinetic term of the Hamiltonian ($\propto \ps^2$).
The possibility to obtain a nonlinear operation $\UU\s{NL}$ in a realistic setup is studied in~\cite{rakhubovsky_stroboscopic_2021,roda-llordes_numerical_2024,roda-llordes_macroscopic_2024,neumeier_fast_2024}.

Below, using the Heisenberg picture, we show that the QND-based unitary transformation illustrated
in~\cref{fig:fig0-pdf}~(a)
which is formally given by the operator
\begin{equation}
  \label{eq:protocol}
  \left( \UU_{qy } (g_{4}) \UU_{px } (g_{3}) \right)
  \cdot
  \UU\s{NL} (\xs, \gamma)
  \cdot
  \left( \UU _{px } (g_{2}) \UU_{qy } (g_{1})\right)
\end{equation}
can induce nonlinear transformations of the target system.
In the equation above, the operators are arranged in the groups corresponding to three steps.
During the first step, the quadratures of the target system are uploaded to the source system in a suitably tailored manner using the linear QND interactions.
In the second step, the nonlinear evolution of the source system implements the nonlinear processing of these uploaded quadratures.
The final step transfers the evolved quadratures back to the target system, which completes the broadcasting of the nonlinearity.

In the Heisenberg picture, the full transformation of the target quadratures induced by the entire protocol reads
\begin{align}
  \label{eq:4qnd_gate}
  \xt\s{i} \mapsto \xt\s{f} &
  = (1+g_4(g_2+g_3))\xt\s{i}
  +(g_4 + g_1 (1 + g_4 (g_2 + g_3)))\xs\s{i},
  \\
  \label{eq:4qnd_gateb}
  \pt\s{i} \mapsto \pt\s{f} &
  =(1 + g_1 (g_2 + g_3))\pt\s{i}
  -(g_2+g_3)\ps\s{i}
  +g_3 \gamma V' \left((1+g_1g_2)\xs\s{i}+g_2\xt\s{i}\right).
\end{align}

By choosing the gains of the Gaussian operations to be $g_1 = - g_4 = - g^{-1}$, $g_2 = - g_3 = g$, the target quadratures are transformed as
\begin{align}
  \label{eq:unitary_broadcast_io}
  \xt\s{i} \mapsto \xt\s{f} & = \xt\s{i},
&
  \pt\s{i} \mapsto \pt\s{f} & =  \pt\s{i} - g \gamma V' ( g \xt\s{i}).
\end{align}
These transformations leave the position of the target $\xt$ unperturbed by the interactions, and the momentum $\pt$ picks a displacement proportional to a nonlinear function $V'( g \xt\s{i})$ of the initial position quadrature $\xt$, dictated by the potential function $V$. 
This displacement is amplified by the gain of the linear QND interactions $g$.
The transformations~\eqref{eq:unitary_broadcast_io} are the same as if we amplified the target in $\xt$, then applied the nonlinear transformation directly to it, and squeezed the target in $\xt$ after the nonlinearity.
Formally, the sequence of the gates applied to the target system directly $\UU\s{S} (g^{-1}) \cdot
\UU\s{NL} (\xt , \gamma ) \cdot
\UU\s{S} (g)$ yields the same transformations of the quadratures ($\UU\s{S}$ is a squeezing operation: $\UU\s{S}^\dag (g) \xt \UU\s{S} (g) = g \xt$).
This amplification is a direct consequence of using the QND interactions for the quadrature exchange between the source and the target: by addressing one quadrature at a time, it is possible to selectively amplify it.
Such an effect would not be directly possible using a beamsplitter-like (state-swap) interaction, which is an advantage of our proposed protocol.

Note that our aim here is to broadcast the nonlinear operation using linear QND operations rather than engineer a certain final state of the target.
It is thus especially important that the transformations~\eqref{eq:unitary_broadcast_io} apply to \emph{an arbitrary} initial state of the target.
This initial state is then transformed in a desired nonlinear fashion, with the effective strength of the nonlinearity enhanced by the gain of the linear QND operations.
Furthermore, the transformations in~\eqref{eq:unitary_broadcast_io} are free from the quadratures of the source system.
Therefore, the initial quantum state of the source does not influence the broadcasting of the nonlinearity.
This property of the broadcasting procedure is particularly beneficial, e.g.\ when reaching an initial pure state of the source is challenging.

The performance of the broadcast nonlinear phase gate on the target (atoms) can first be evaluated at the level of the first and second statistical moments of quadratures.
The input-output relations~\cref{eq:unitary_broadcast_io} suggest that the protocol does not influence the statistics of the amplitude quadrature $\xt$ of the target.
The mean value and the variance of the momentum quadrature (as well as the higher-order moments), however, change by values proportional to the nonlinearity $\gamma$, and the strength of the linear gate $g$.
These changes of the first and second moments can be examined already at the classical level.

Furthermore, the nonlinearity on the atoms can be evaluated using the \emph{nonlinear variance}~(NLV)~$\nlv$ and compared to the nonlinear phase gate on the source (mechanics)~\cite{moore_estimation_2019,rakhubovsky_stroboscopic_2021}.
Being a property of a quantum state $\rho$, the NLV corresponding to the nonlinear potential $V$ is defined as
\begin{equation}
  \label{eq:nlv_definition}
  \nlv (\lambda) = \Var_\rho ( \pt - \lambda V'( \xt ) ),
\end{equation}
where $\Var_\rho(\cdot)$ denotes the variance evaluated with respect to the quantum state $\rho$.
The function $\nlv(\cdot)$ is, by construction, a quadratic function of the real-valued parameter $\lambda$.
Application of a nonlinear phase gate $\exp[ - \ii \gamma V (\xt) / 2]$ to a state of a harmonic oscillator creates correlations between nonlinear combinations of the quadratures.
As a result, the value of the NLV is suppressed for certain values of $\lambda$ below the level corresponding to its initial quantum state.
The value of $\lambda$ at which~$\nlv(\lambda)$ reaches its minimal value indicates the strength of the nonlinear correlations of the examined state.
The reverse is also true: suppression of the NLV indicates the existence of the correlations between the nonlinear combinations of the quadratures peculiar to a certain type of nonlinearity~\cite{moore_hierarchy_2022}.
Moreover, there are bounds for the values of the NLV that apply to all classical states and all Gaussian states~\cite{moore_hierarchy_2022}, so that violation of the corresponding threshold by a certain state indicates non-classicality or quantum non-Gaussianity of this state.
The graphs of the NLV corresponding to the cubic potential $V(x) \propto x^3$ for a few example quantum states are in~\cref{fig:fig0-pdf}~(d).
The properties of the NLV and analogies with the evaluation of Gaussian squeezing are further discussed in~\cref{sec:nonlinear_variance_as_a_figure_of_merit}.

For the state of the target after the transformations~\eqref{eq:unitary_broadcast_io}, the NLV reads (assuming $\ev{\xt\s{i}} = \ev{ \pt\s{i}} = \ev{ \{ \xt\s{i} , \pt\s{i} \} } = 0$):
\begin{equation}
  \label{eq:nlv_bcast_unitary}
  \nlv (\lambda)
  =
  \Var \pt\s{i} + \Var \Big( g \gamma V' (g \xt\s{i}) - \lambda V' ( \xt\s{i} ) \Big)
  =
  \Var \pt\s{i} + ( \lambda - \gamma g^n )^2 \Var( V' ( \xt\s{i}) ),
\end{equation}
where the second equality is written for the case of a monomial potential $V(x) = x^n / n$.
The NLV is suppressed for the values around $\lambda = \gamma g^n$.
The values at which the suppression can be observed indicate the strength of the nonlinearity required to reach the same suppression by applying the nonlinear gate of the same kind (with same $V(x)$) to the ground state.
The suppression of the NLV that we observe in the target system is directly equivalent to the application of the nonlinearity to the target system.
The minimal value of the NLV allows estimating the amount of noise added during the nonlinear operation.
In the case of~\cref{eq:nlv_bcast_unitary} it is bound from below by the initial variance of the target's momentum quadrature $\Var \pt\s{i}$.
This is equivalent to an action of a unitary nonlinear gate.

The full broadcasting protocol (\cref{fig:fig0-pdf}~(a)) uses four unitary linear QND interactions to transfer the nonlinearity regardless of the initial quantum state of the source.
Since, in a realistic setup, both the source and target systems are open, the Gaussian operations inevitably introduce environmental noise, and unitary interactions are only an approximation of the realistic dynamics.
The performance of a realistic setup is further investigated in~\cref{sec:implementation}.
To explore the case when approaching unitary linear interactions is challenging for the initial experimental implementation, we consider a \emph{simplified protocol} with only two linear interactions illustrated in~\cref{fig:fig0-pdf}~(b).
This scheme uses one linear QND gate before and one after the nonlinear transformation, which is formally equivalent to setting $g_1 = g_4 = 0$ in~\cref{eq:4qnd_gate,eq:4qnd_gateb}.
Assuming equal gains of the remaining linear interactions $g_2=-g_3=g$, we obtain the following input-output relations for the target system:
\begin{align}
  \label{eq:two_qnd_unitary}
  & \xt\s{f}=\xt\s{i},
  &
  & \pt\s{f}=\pt\s{i}- g \gamma V'\left(\xs\s{i}+g\xt\s{i}\right).
\end{align}

The two-QND-gate scheme above (\cref{fig:fig0-pdf}~(b)) aims to approach the application of the nonlinear transformation to the target system if the noise of $\xs\s{i}$ is sufficiently suppressed.
Alternatively, the full scheme with four QND interactions can instead be optimized for a stronger suppression of the NLV of the target system by a different choice of \gains.
Consider the transformations provided by $g_2 = g - 1/g_1$, $g_3 = - g$, $g_4 = 1/g$.
After these non-unity gain transformations, the target system has quadratures given by
\begin{align}
  \label{eq:target_squeez}
  \xt\s{f}& =
  \left(1-\frac{1}{g g_1}\right) \xt\s{i}
  + g_1\xs\s{i},
  &
  \pt\s{f}& =\frac{1}{g_1}\ps\s{i}
  -g \gamma V' \left( g \xt\s{f}\right).
\end{align}

For a monomial potential $V(x) = x^n /n$, the corresponding NLV takes the form similar to~\cref{eq:nlv_bcast_unitary}:
\begin{align}
  \label{eq:nlv_sqzgen_unitary}
  & \nlv ( \lambda ) =  \frac{1}{g_1^2} \text{Var}(\ps\s{i}) +  (\lambda- \gamma g^n)^2\text{Var}(V'(\xt\s{f}))
\end{align}
The NLV in this regime is again suppressed for the values around $\lambda = \gamma g^n$.
In contrast to the \emph{broadcasting} regime, here the NLV can be made arbitrarily close to zero by choosing a stronger gain $g_1$.

Moreover, the concluding QND interaction of each scheme can, in fact, be substituted by an optical readout of the momentum quadrature of the source system $\ps$ (with the outcome $\bar p$) followed by a coherent displacement of the momentum quadrature of the target system $\pt$:
\begin{equation}
  \pt \mapsto \pt + K \bar p,
\end{equation}
where $K$ is the coefficient set during the coherent displacement.

To summarize the above, the proposed atom-mechanical scheme can operate in two regimes.
The first is the \emph{nonlinearity broadcasting regime}.
In this regime, the final state of the target is as if the nonlinear transformation has been applied directly to this mode, see~\cref{eq:unitary_broadcast_io}.
We propose two ways of broadcasting.
The first way uses four linear QND interactions between the source and the target and is capable of reaching a perfect broadcasting given the unitary dynamics of the system.
The second way uses only two linear QND interactions and may be more practical where reaching unitary QND interactions is challenging.
Importantly, both regimes allow us to not only broadcast the nonlinearity but also to amplify it by virtue of the strength of the linear QND interactions.
The second proposed non-unity gain regime is the \emph{nonlinear-squeezing generation regime}.
In this preliminary regime, instead of effectively applying the nonlinear transformation to the target mode, its nonlinear squeezing is maximized, see~\cref{eq:target_squeez}.
In the following sections, we discuss the specific physical implementation of the protocol using an atom-mechanical system and suggest the configurations to reach the optimal performance of the discussed regimes.

\subsection{Implementation of nonlinearity broadcasting from levitated nanoparticle to atomic cloud} \label{sec:implementation}

In this section, we discuss an attractive example of the implementation of the protocol in a concrete hybrid system.
We consider an atomic ensemble
with collective spin variables equivalent to a linear harmonic oscillator
as a target, and a nonlinear mechanical oscillator of an optomechanical cavity as a source.
We outline the details of the implementation of a linear QND interaction and a nonlinear evolution of the source in such a system, and then apply these implementations to analyze the full broadcasting protocol.

Spin ensembles at room temperature are well known for the capabilities for storage of continuous-variable quantum information~\cite{cerf_quantum_2007}, and they lack a native nonlinear transformation, which naturally suggests them for the target.
The mechanical oscillator can be a source of nonlinearity thanks to its intrinsic nonlinear rigidity~\cite{ochs_amplification_2021,ochs_frequency_2022} or due to an engineered nonlinear potential as in the case of levitated nanoparticles~\cite{siler_diffusing_2018,gonzalez-ballestero_levitodynamics_2021,rakhubovsky_stroboscopic_2021,neumeier_fast_2024,roda-llordes_macroscopic_2024}.
The way to establish a linear QND interaction in such a hybrid system is described in detail in~\cite{manukhova_pulsed_2020}.
In brief, a pulse of mediating squeezed light (duration $\tau$, squeezing $S$) is first made to interact sequentially with both atoms and mechanics, accompanied by classical drive.
The local interactions are engineered to be of a linear QND type with the interaction rates $g\s{a}$ and $g\s{m}$.
We assume each system (atoms and mechanics) to be in cavities of equal linewidths ($\kappa\s{a} = \kappa\s{m} =\kappa$).
After the pulsed interactions, a quadrature of the optical mediator is detected using a homodyne detector, and a magnetic feedforward displaces the atomic ensemble based on the measurement outcome.
The input-output transformations that describe such a QND gate between atoms and mechanics read
\begin{align}
  \label{eq:real_qnd}
  \xt\up{out}
  & =\eta_X \xt\up{in}
  +\hat{N}_{X},
&
  \xs\up{out}
  & =\eta_q \xs\up{in}
  -  \cG\s{m} \xt\up{in}
  +\hat{N}_{q},
\\
  \pt\up{out}
  & =\eta_Y \pt\up{in}
  +\cG\s{a} \ps\up{in}
  +\hat{N}_{Y},
&
  \ps\up{out}
  & =\eta_p \ps\up{in}
  +\hat{N}_{p},
\label{eq:real_qnd_a}
\end{align}
where $(\xt,\pt)$ are the quadratures of the atomic subsystem, $(\xs,\ps)$ correspond to the mechanical subsystem, and the operators with the superscript $\sbullet\up{in/out}$ denote the corresponding operators before/after the interaction.
We assume the atoms to be initially in the ground state, and the mechanics in a squeezed thermal state with mean occupation $\bar n$ and squeezing $S\s{m}$ (the initial variance of the position is reduced by a factor $S\s{m}$).
The quantities $\eta_{\sbullet} \approx 1$ embody the damping in both subsystems, which is very low thanks to the high quality of both systems.
E.g., in Ref.~\cite{thomas_entanglement_2020}, the intrinsic quality factors of atomic and mechanical oscillators are, respectively, $\mathcal Q\s{a} \approx \num{1e3}$ and $\mathcal Q\s{m} \approx \num{7e8}$.
For optically levitated dielectric nanoparticles, Ref.~\cite{dania_ultrahigh_2024} reported $\mathcal Q\s{m} > 10^{10}$.

Here we write the input-output relations for the interaction designed to mimic the Hamiltonian $\propto \xt \ps$.
The other required interaction ($\propto \pt \xs$) can be equivalently achieved by changing the phases of the classical drives.
In particular, for the mechanics, the local interaction takes the form~\cite{braginsky_quantum_1980}
\begin{equation}
  \label{eq:optomech-qnd-coupling}
  H\s{mL} = g\s m ( \xs \cos \theta\s m + \ps \sin \theta \s m )( \hat X\s L \cos \theta\s{L} + \hat P\s L \sin \theta \s L),
\end{equation}
where $(\hat X\s L, \hat P\s L)$ are the quadratures of the mediating light.
The phase $\theta\s m$ can be controlled by the phase of the classical drive, and the phase $\theta\s L$, by a local phase-rotation operation on the light.
Control over these two phases allows engineering different optomechanical coupling Hamiltonians, e.g., $\propto \xs \hat P\s L$ or $\propto \ps \hat X\s L$.
Combined with a similar possibility on the atoms, we can construct both required atom-mechanical QND couplings.

The parameters $\cG\s{a}$ and $\cG\s{m}$ above represent the effective gains of the linear transformation, defined by the parameters of the local atom-optical and optomechanical interactions, feedforward strength, and the optical loss in the mediator.
These coefficients can be made equal ($\cG\s a = \cG \s m = \cG$), in which case the hybrid atom-mechanical gate corresponds, up to the admixed noises, to a Hamiltonian transformation with the evolution operator $\exp[ - \ii \cG \xt \ps / 2]$.
This fundamentally important regime (termed \emph{gain-symmetric}), however, might not be optimal for the broadcasting of the nonlinearity.
Therefore, we separately consider the regime where the two gains $\cG\s{a,m}$ can be unequal (we refer to it as \emph{gain-asymmetric}).
In realistic setups (e.g.~\cite{thomas_entanglement_2020}), values $\cG\s{a,m} \sim 50$ are reachable~\cite{manukhova_pulsed_2020}.

The noise operators $\hat N_{\sbullet}$ originate from the environments of the atomic and mechanical subsystems, and from the losses of the optical mediator.
A substantial contribution to the noise is from the mechanical environment, characterized by the mechanical heating rate $\heating = \damping\s{m} ( n\s{th} + 1/2 )$, where $\damping\s{m}$ is the damping rate of the mechanical oscillator and $n\s{th}$ is the mean thermal occupation of the mechanical bath.
Moreover, when the optical mediator is not fully decoupled from the source-target system, its initial squeezed state also contributes to noise.

To realize the nonlinear broadcasting gate, in addition to QND interactions, we essentially need a suitable nonlinear evolution of the target, i.e. of the mechanics.
One of the promising mechanisms of a versatile implementation of the nonlinearity is to use the center of mass motion of a levitated nanoparticle in a potential beyond quadratic~\cite{siler_diffusing_2018,gonzalez-ballestero_levitodynamics_2021,rakhubovsky_stroboscopic_2021,neumeier_fast_2024}.
As a dielectric in an inhomogeneous optical beam, the nanoparticle experiences a force toward the stronger field intensity.
Hence, the effective potential for the mechanical motion is proportional to the spatial intensity profile of the trapping beam.
The latter can be engineered to be a nonlinear function of the particle's displacement beyond the quadratic potential of a harmonic oscillator.

We consider the nonlinear evolution very short, with its duration being of the order of a few periods of mechanical oscillation.
Thanks to the high $Q$-factor of the mechanical oscillator, this enables us to ignore the interaction of each system with the environment and consider this step purely unitary with good accuracy.
For a sufficiently heavy particle, we can neglect the free-motion kinetic term $\ps ^2 / ( 2 m )$ in the mechanical motion Hamiltonian ($m$ is the effective mass of the particle).
Such dynamics can be achieved via pulsed stroboscopic manipulation of the nonlinear potential~\cite{rakhubovsky_stroboscopic_2021,neumeier_fast_2024} or employing a nanoparticle with sufficiently large mass $m$ to suppress the kinetic term.
For our demonstration, we assume a cubic potential $V(x) \propto x^3$ of the mechanical motion.

Using the four atom-mechanical gates described by the equations of the type~\eqref{eq:real_qnd}, each with potentially different effective gains $\cG_i$,
and the cubic phase gate on the mechanics,
it is possible to approximate different protocols of unitary dynamics, outlined in the previous section.
Moreover, in principle, in each of the protocols, it is not necessary to implement the full atom-mechanical QND gate on the last step.
The role of this last interaction is solely to upload the quadrature $\ps$ to the atoms.
Therefore, the atom-mechanical QND interaction can eventually be substituted by an optomechanical detection of the mechanical momentum $\ps$, and a subsequent feedforward displacement of the atomic quadrature $\pt$ with the result of the detection.

In a realistic setup, the benefits of the broadcasting protocol can be accompanied by adding extra noise to the system.
In the following section, we, therefore, evaluate the performance of the proposed nonlinear atom-mechanical QND protocols with relevant imperfections.
We demonstrate that a system within experimental reach can effectively broadcast the nonlinearity from the mechanics (source system) to the atoms (target system) despite the acceptable imperfections.

\subsection{Evaluation of the broadcast nonlinearity}  \label{sec:evaluation}

In this section, we present the simulations of the broadcasting schemes proposed in~\cref{sec:nonlinearity_distribution_to_a_linear_system_4QND} applied to the particular hybrid system described in~\cref{sec:implementation}.
The simulations are performed symbolically using techniques outlined in~\cref{sec:methods}.

\begin{figure}[htb!] \centering
  \begin{tikzpicture}[plt/.style={anchor=north west},
    labl/.style = {},
    ]
    \node [plt] (a) {\includegraphics[width=.49\linewidth]{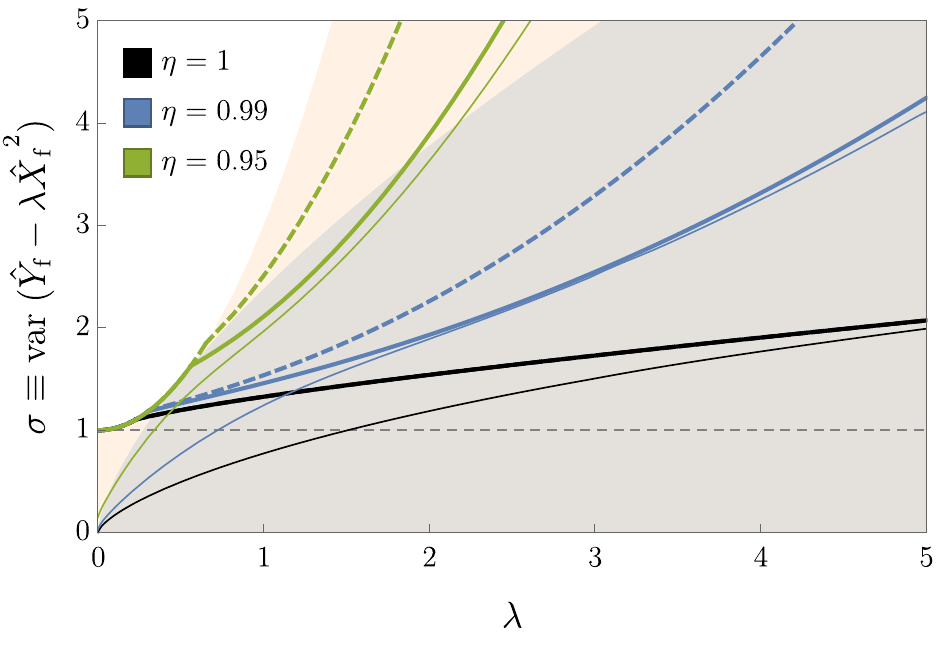}};
    \node [plt] at (a.north east) (b) {\includegraphics[width=.49\linewidth]{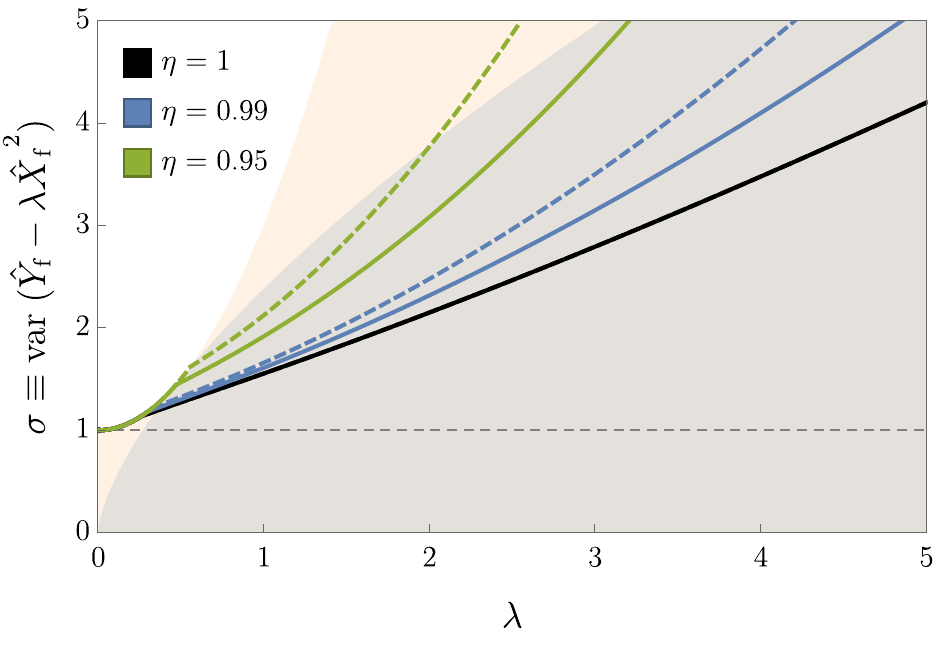}} ;

    \node [anchor = north west] (la) at (a.north west) {\sffamily\bfseries (a)} ;
    \node [anchor = north west] (lb) at (b.north west) {\sffamily\bfseries (b)} ;

    \node [anchor = south east] at ($(a.south east)+(0,1cm)$) {\includegraphics[page=1, scale=0.65]{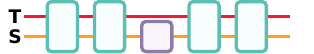}} ;
    \node [anchor = south east] at ($(b.south east)+(0,1cm)$) {\includegraphics[page=2, scale=0.65]{./tikz/fig_inset_circuit.pdf}} ;

    \node at (a.north) {\sffamily Full Protocol};
    \node at (b.north) {\sffamily Simplified Protocol};
  \end{tikzpicture}\caption{
    Nonlinearity broadcasting to the atomic ensemble, visualized by the envelopes of the nonlinear variance $\nlv (\lambda)$ curves.
    The nonlinear variance [see~\cref{eq:nlv_definition}] of the final state of the atomic ensemble after
    (a)~the full protocol,
    (b)~the simplified protocol, see~\cref{fig:fig0-pdf}.
    Areas reachable only by quantum non-Gaussian and non-classical states are indicated by, respectively, gray and light orange shading.
    In~(a), thick lines show results of the protocol in the \emph{broadcasting} regime.
    Dashed lines correspond to~\emph{gain-symmetric} QND gates, full lines to~\emph{gain-asymmetric}.
    Different colors correspond to different loss of the mediator~$1 - \eta$.
    In the lossless case $\eta = 1$, both gain-symmetric and gain-asymmetric regimes correspond to the same configuration and hence perform identically.
    Thin lines show the optimized performance of the \emph{nonlinear squeezing} generating regime.
    For all lines of this panel, the numerical parameters are: nonlinearity $\gamma = 0.07$, heating of the mechanics $\heating = 10^{-4}\kappa$.
    The mechanics is initially in a thermal state with mean occupation $n = 0.45$ without squeezing.
    In~(b), full and dashed lines, similarly, correspond to using gain-asymmetric and gain-symmetric QND gates in the simplified protocol.
    The mechanical oscillator is initially in a squeezed thermal state with mean occupation (before squeezing) $n = 0.45$, squeezing $7$ dB.
  }
  \label{fig:2&4qnd_nlv_real}
\end{figure}

For the initial evaluation of the protocol's performance, one can use the first and second statistical moments of the quadratures of the atoms.
Indeed, thanks to the nonlinearity, already a zero-mean initial state of the atoms acquires a non-zero shift of the mean value of the momentum quadrature~$\pt$, as follows from~\cref{eq:unitary_broadcast_io}.
To evaluate the correlations in the quantum noise created by the nonlinearity broadcasting, we use the suppression of the nonlinear variance (NLV) $\nlv$.
For the cubic nonlinearity $V(x) = x^3 / 3$, which we consider here, the relevant NLV over the final state of the atomic system can be computed as
\begin{equation}
  \label{eq:nonlin_var_def}
  \nlv (\lambda) = \Var ( \pt\s{f} - \lambda \xt\s{f}^2 ).
\end{equation}
We compare this quantity with two important thresholds, first $\nlv\s{NC}$ that demarcates the values reachable by classical states (coherent states and convex mixtures thereof) from the ones reachable only by the non-classical states.
The second one $\nlv\s{NG}$ similarly distinguishes convex mixtures of Gaussian states (squeezed displaced vacua) from genuinely \emph{quantum non-Gaussian} states.
These thresholds are given by~\cite{moore_hierarchy_2022,kala_cubic_2022}
\begin{align}
  \label{eq:nonlin_var_thrs}
  & \nlv\s{NC} (\lambda) = 1+2\lambda^2,
  &
  & \nlv\s{NG} (\lambda) = 3 \sqrt[3]{ \lambda^2 / 2 }.
\end{align}
Whenever the NLV, evaluated for a state $\rho$, is such that for at least one $\lambda'$, $\nlv_\rho (\lambda') < \nlv\s{NC} (\lambda')$, the state $\rho$ cannot be represented as a convex mixture of coherent states and, therefore, is nonclassical.
Similarly, if $\nlv_\rho( \lambda') < \nlv\s{NG} (\lambda')$ for some $\lambda'$, the state $\rho$ cannot be represented as a convex mixture of Gaussian states and is quantum non-Gaussian.
The quantum non-Gaussianity threshold is formed by NLV evaluated for optimal Gaussian states.
For low values of $\lambda$, where the dominant contribution to NLV is given by the momentum, the optimal Gaussian states are the momentum-squeezed states.
For higher values of $\lambda$, the optimal Gaussian states are position-squeezed states.
At $\lambda = 1/2$, the optimal Gaussian state is the ground state, whose NLV also gives the non-classicality threshold, and the two thresholds touch.
In all figures, the corresponding areas below the thresholds are shaded.
Another important line is obtained by considering the unitary application of a nonlinearity directly to the target system (equivalent to the ideal performance of the broadcasting protocol) as in~\cref{eq:unitary_broadcast_io}.
Given an initial ground state of the target, such an application yields the NLV given by $\nlv (\lambda) = 1 + 2 ( \lambda - \mu )^2$ for each value of the effective nonlinear parameter $\mu = \gamma g^3$.
Sweeping through different values of $\mu$ continuously, the ideal protocol yields a family of parabolas, whose collective envelope is a horizontal line at the level of the ground-state momentum variance $\nlv = 1$ due to the assumption of the initial ground state of the target
(thin dashed gray lines in~\cref{fig:2&4qnd_nlv_real}).

We start with the full nonlinear protocol implemented with all four linear QND gates (\cref{fig:fig0-pdf}~(a)) in the broadcasting regime.
Using parameters inspired by the recent experiments~\cite{delic_cooling_2020,thomas_entanglement_2020}, we study the protocol with the linear QND gains suggested for the unitary broadcasting (values before~\cref{eq:unitary_broadcast_io}).
To estimate the optimal performance of a realistic setup, we first fix the initial states of the atoms (ground state) and the mechanics (squeezed thermal state), the nonlinearity of the mechanics $\gamma$, the heating rate $\heating$ and the loss of the mediator $1 - \eta$.
Subsequently, we vary the Gaussian interaction gains $-10 \leq \cG_i \leq 10$ (maintaining their relationship as dictated by the broadcasting regime [see~\cref{eq:unitary_broadcast_io}]) and the mediator squeezing $S$, and consider the families of the obtained NLV curves.
We then minimize the NLV over the variables $\cG_i$ and $S$ for each value of~$\lambda$.
The resulting values of the minimized NLV as a function of $\lambda$ form the \emph{envelope} of the NLV curves.
The optimization is required because, in the picture that includes realistic sources of the noise acting on the mechanics, the noise entering the system at earlier stages of the protocol can be amplified during the subsequent stages.
As a result, the simple strategies suggested in the unitary picture do not hold, and have to be substituted by the optimization.
For each certain parabola of NLV, the position of its minimum indicates the effective nonlinearity strength, and the minimal value allow one to assess the quality of the nonlinearity.
The envelopes that we obtain for each regime, therefore, visualize the relationship between the attainable strength and quality of the nonlinearity.
Further details of the optimization procedure are in~\cref{sec:Variance_minimization}.

\cref{fig:2&4qnd_nlv_real}~(a) demonstrates the resulting envelopes of the curves of the NLVs.
Dashed lines correspond to the \emph{gain-symmetric} regime (discussed in~\cref{sec:implementation}) where the hybrid Gaussian atom-mechanical gates are constructed to mimic the Hamiltonian evolution up to the noises.
Full lines correspond to the \emph{gain-asymmetric} regime, so that each gate is characterized by two gains of linear interactions ($\cG\s{a}{}_i$ and $\cG\s{m}{}_i$).
The relation between these two gains in each Gaussian gate is maintained in a way that suppresses the contribution of the optical mediator in the final quantum state of the atoms (see~\cref{sec:Variance_minimization}).
The gain-asymmetric regime shows a visibly better performance (lower values of the NLV) due to this suppression.
Importantly, both regimes are capable of reaching non-classical and quantum non-Gaussian states of the atoms via the nonlinearity broadcasting, given sufficiently low loss of the mediator.
Note that in the lossless case~($\eta = 1$), the gain-asymmetric regime is automatically reduced to the gain-symmetric regime, therefore their performance is the same.

Compared to the predictions of the unitary regime (\cref{eq:nlv_bcast_unitary,eq:nlv_sqzgen_unitary}), the optimized performance of both regimes is limited by the optical loss and mechanical noise.
The noise contributions to the final state of the target do not possess the nonlinear correlations required for the suppression of the NLV.
As a consequence, the minima of the NLV parabolas become elevated with respect to the performance of the unitary protocols.
In particular, this results in the minimal $\lambda$ for which the non-classicality or quantum non-Gaussianity can be observed.
On the graphs, for some values of $\lambda$, the NLV envelopes coincide with the non-classicality threshold.
For these values of $\lambda$, broadcasting the nonlinearity to atoms yields the atomic quantum state with higher NLV than the initial vacuum state.
The primary noise sources in our work are the loss of the optical mediator~$1 - \eta$ and the heating of the mechanical oscillator~$\heating$.
The experiment~\cite{delic_cooling_2020} reported the heating rate~$\heating$ bounded from below by the recoil heating rate $\heating \approx 0.06 \kappa$.
This value can be further suppressed by using a longer wavelength of trapping laser, by applying additional squeezed light to minimize the recoil scattering~\cite{gonzalez-ballestero_suppressing_2023}, or trapping in an intensity minimum~\cite{dago_stabilizing_2024,mlynar_feedback_2025}.
For the calculations, we used $\heating \leq 10^{-4} \kappa$.
Another source of noise in NLV is the non-zero initial mechanical occupation, which can be suppressed by optical cooling~(e.g., see~\cite{delic_cooling_2020,kamba_optical_2022,piotrowski_simultaneous_2023}).
For simulations, we used $\bar n=0.45$; however, values of the initial occupation below $\bar n \leq 10$ have virtually no influence on the performance of the broadcasting protocol.
Additionally, the optical loss of the mediator contributes to an increase in the NLV.
As visualized by the figure, even a loss of a few percent can be critical and cause the loss of quantum non-Gaussianity, even when a careful optimization of the gains has been performed.
Lastly, contributions of the optical mediator to the final state of the atoms cause an increase of the NLV.
For the protocol with \emph{gain-symmetric} linear QND gates, there exist optimal values of the mediator squeezing of approximately $9.5$~dB that help to reduce these contributions.
For the gain-asymmetric gates, the higher the squeezing, the stronger the established nonlinear correlations~(the lower the NLV curves).

In~\cref{fig:2&4qnd_nlv_real}~(b), we present the results of the simulations of the simplified protocol with only two linear QND gates.
By inspection, we see that the optical loss of the mediator has a relatively weaker impact on the protocol performance.
This is simply due to the fewer number of interactions of the matter modes with the mediator.
The performance of the protocols based on gain-symmetric and gain-asymmetric gates differs less in this scenario compared to the full protocol with four QND gates for the same reason.
The requirements to the initial cooling of the mechanical oscillator are more stringent in the simplified protocol, as dictated by~\cref{eq:two_qnd_unitary}, where the quadratures of the initial state of the mechanics are not fully eliminated from the final state of the atoms.
However, the noise in the two quadratures of the initial mechanical state contributes differently to the final state of the atoms.
It is possible to take advantage of this by pre-squeezing the motion of the nanoparticle~(e.g.,~\cite{bonvin_state_2023,duchan_nanomechanical_2025,rossi_quantum_2024,tomassi_accelerated_2025} for the levitated NPs).
Other techniques to achieve the squeezing of mechanical motion have been proposed in~\cite{mari_gently_2009,rakhubovsky_squeezing_2013,neumeier_fast_2024,kustura_mechanical_2022}.
Furthermore, it follows from~\cref{eq:two_qnd_unitary} that the influence of the initial mechanical state can be compensated by using a weaker nonlinearity $\gamma$ and stronger Gaussian gains $g$.
In~\cref{fig:2&4qnd_nlv_real}~(b), we assumed the initial mechanical state to be a thermal squeezed state with initial occupation $n = 0.45$ before the squeezing and $7$~dB of squeezing.
The positive role of the mechanical squeezing gets saturated at this value, so that using a stronger squeezing does not improve the NLV, because other sources of noise become dominant.

The two protocols show a different dependence on the nonlinear parameter $\gamma$.
For the full protocol, higher values of $\gamma$ are beneficial for the suppression of the NLV.
Surprisingly, for the simplified case, increasing $\gamma$ might cause an increase of the NLV.
This is because in a realistic scheme with noise, the admixed noise is also amplified by the nonlinear element; therefore, for a stronger nonlinearity, there is a larger impact of the noise.

\begin{figure}[htb!]\centering
\includegraphics[width = .95\linewidth]{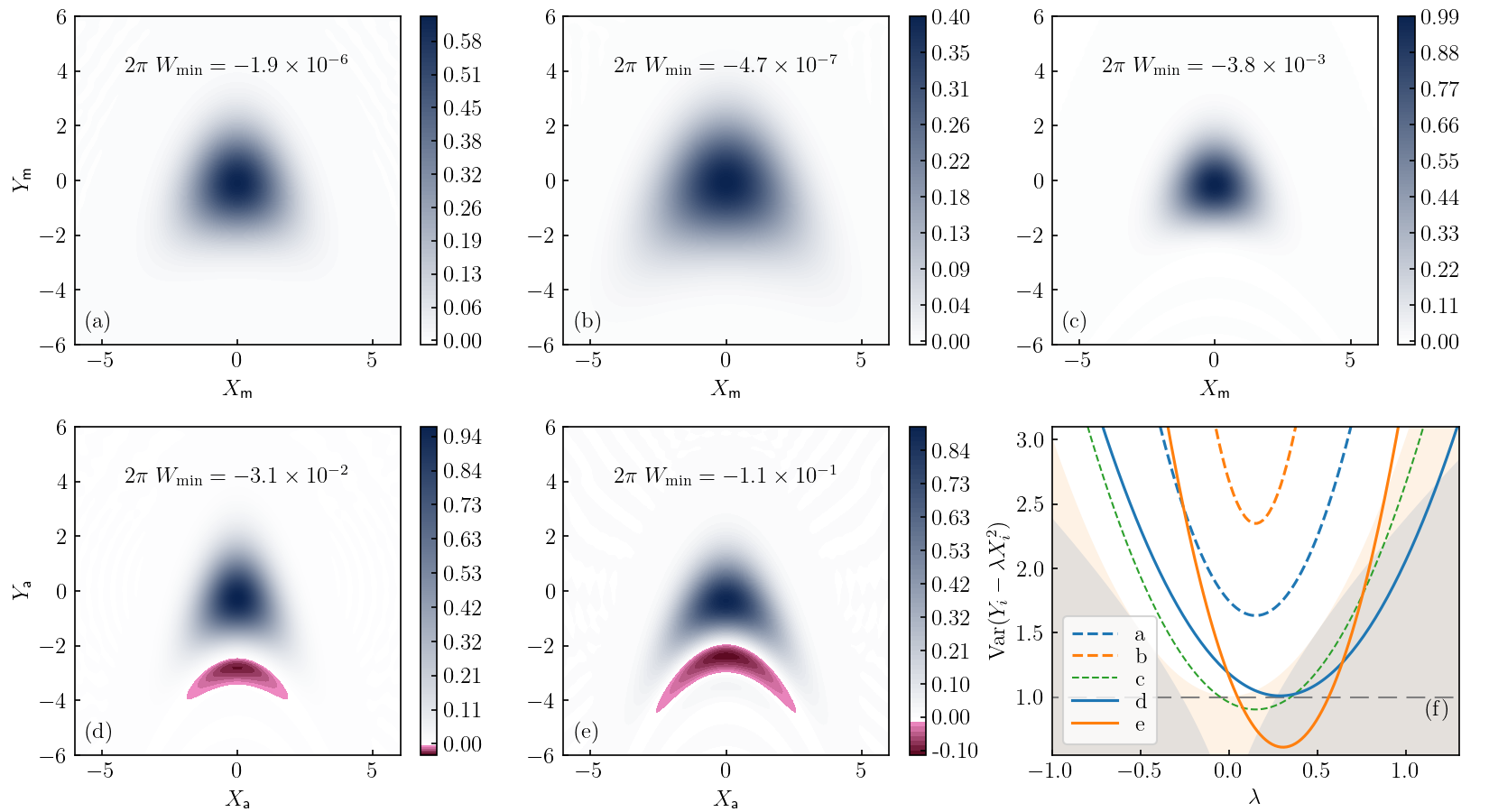}
    \caption{Broadcasting of the nonlinearity witnessed by the Wigner function negativity and the nonlinear variance suppression.
        Wigner functions $2 \pi \; W (X,Y)$ of (a,b) the mechanical oscillator at $t = t\s{NL}$ (after the application of the nonlinearity) and (d, e) of the atomic system at the end of the protocol ($t = t\s{f}$).
        The graphs are made in the regime of (a, d) broadcasting the nonlinearity, (b, e) generation of nonlinear squeezing.
        In (c) the nonlinearity of the same strength is applied directly to the initial state of the mechanics.
        (f)~Nonlinear variance $\nlv(\lambda)$ evaluated for the corresponding state.
        Atoms are initially in vacuum, mechanics in a squeezed vacuum state with squeezing $r =0.1$.
        Numerical parameters:
        nonlinearity $\gamma = 0.15$,
        atomic decay rate $\zeta\s a = 10^{-3}$,
        mechanical decay rate $\zeta\s m = 10^{-6}$,
        mechanical bath occupation $n\s{th} = 10^3$, which results in the heating rate $\heating = 10^{-3}$.
        The values defining the QND gains (see~\cref{eq:unitary_broadcast_io,eq:target_squeez}) are $g = 1.26$ and $g_1 = 1.2$.
    }
    \label{fig:wf}
\end{figure}

In~\cref{fig:wf} we present the results of the numerical simulation of the system in two regimes: broadcasting of the nonlinearity and the generation of the nonlinear squeezing.
For each regime, here we solve numerically the corresponding Lindblad master equation (see~\cref{sec:numerical_lindblad}) for an open-system dynamics of the \emph{bipartite} atom-mechanical system.
Due to technical reasons, we are using idealized parameters (reduced heating and initial occupation of the mechanical oscillator), in order to prove the principal possibility of the protocol and illustrate the highly nonclassical and nonlinear attainable quantum states of the atomic system.
In this section, we consider the variants of the full protocol aimed at broadcasting the nonlinearity, and at the generation of nonlinear correlations (\cref{fig:fig0-pdf}~(a)).

A common theoretical way to verify the non-classicality of a quantum state is to observe negative values of the Wigner function~\cite{weinbub_recent_2018}.
In particular, the negativity of the Wigner function (WF) is a direct witness of quantum non-Gaussianity~\cite{walschaers_nongaussian_2021,lachman_quantum_2022,rakhubovsky_quantum_2024a}.
In~\cref{fig:wf}~(a,b) we present the WFs corresponding to the state of the mechanical mode right after the nonlinear transformation is applied to it.
These states already have distinctively non-Gaussian shapes of the Wigner functions; however, the Wigner functions are everywhere positive up to the precision of the numerical simulations.
The quantum states of the atoms are in~\cref{fig:wf}~(d,e).
These quantum states exhibit both visibly non-Gaussian shapes and negative values.
It is interesting that despite the last steps of the protocol involving only Gaussian operations, the WFs of atoms exhibit negative values, as opposed to the mechanical states.
This can be explained as follows.
In the broadcasting regime, before the application of the nonlinearity (at $t = t_2$, see~\cref{fig:fig0-pdf}), the quadratures of the system read
\begin{align}
  \xt_2 & = \xt\s i - \frac 1 g \xs \s i,
        &
  \pt_2 & = - g \ps \s i,
        &
\xs_2 & = g \xt \s i,
        &
  \ps_2 & = \ps \s i + \frac 1 g  \pt \s i.
\end{align}
The mechanical momentum $\ps_2$ contains contributions from both the initial mechanical momentum $\ps\s i$ and the initial momentum of the atoms $\pt\s i$.
The application of the nonlinearity creates the necessary correlations between the mechanical quadratures; however, due to the contribution from $\ps\s i$, these correlations are insufficient for the Wigner negativity to emerge.
The two subsequent linear QND gates transfer these correlations back to the atoms, amplifying them along the way, and, moreover, cancel the term proportional to~$\ps\s i$.
Therefore, the nonlinear correlations in the quadratures of the atoms are much more pronounced.
To emphasize the role of the amplification by linear interactions, we also show the result of applying the nonlinearity of the same strength directly to the mechanical mode in~\cref{fig:wf}~(c).
The resulting Wigner function exhibits a characteristic non-Gaussian shape peculiar to the states resulting from a cubic phase gate, but has only very shallow negativity.
In contrast, the nonlinearity that is broadcast to atoms~(\cref{fig:wf}~(d)) yields an order of magnitude stronger negative values of the Wigner function thanks to the amplification by the linear QND interactions.
Note, however, that the quantum non-Gaussianity of the state in~\cref{fig:wf}~(c) can be confidently witnessed from the NLV.
In~\cref{fig:wf}~(f), the corresponding [densely dashed green] line is well within the region of QNG states for $0.3 \leq \lambda \leq 0.8$.

In each of the illustrated regimes, the properties of the final quantum state of the atoms agree well with the theoretical predictions.
The WFs have the distinctive shape peculiar to a cubic transform applied to the initial thermal state.
They also exhibit characteristic oscillations reaching negative values, a strong proof of non-classicality.
The graphs of the NLVs~(\cref{fig:wf}~(e)) also correspond to the theoretical predictions.
In the regime of broadcasting, the curve of the NLV corresponding to the final state of the atoms is displaced exactly as if the nonlinearity was applied directly to the atoms.
Moreover, it is displaced by the amount $\lambda\s{B} = \gamma g^3 \approx 0.3$ that agrees with~\cref{eq:nlv_bcast_unitary}.
In the regime of nonlinear squeezing generation, the position of the minimum of the parabola is the same. 
Its minimum value equals $0.7 \approx g_1^{-2}$.
Both position of the minimum and the minimal value are in good agreement with the theoretical predictions of~\cref{eq:nlv_sqzgen_unitary}.

\section{Discussion and Conclusion} \label{sec:discussion}

Let us analyze how individual parameters of the experimental realization of the protocol influence its overall performance.
Practically, the performance is characterized by the envelopes of the NLV curves as functions of the parameter $\lambda$.

First, in order to improve the performance of the gate, the decoherence parameters have to be suppressed, as their sole role is the disruptive addition of the noise.
This involves suppression of optical loss by engineering better optical coupling and reducing the environmental mechanical heating of the source system.
If a bulk mechanical oscillator is used as the source, this involves cooling its environment to cryogenic temperatures and reducing the coupling to the environment~\cite{hoj_ultracoherent_2021,hoj_ultracoherent_2024,bereyhi_perimeter_2022,beccari_strained_2022}.
For the levitated nanoparticles, suppression of decoherence requires reaching a nearly perfect vacuum in the trapping chamber and decreasing the recoil heating.
The latter is possible by decreasing the trapping power, increasing the nanoparticle's mass, or using more elaborate optical techniques for trapping~\cite{gonzalez-ballestero_suppressing_2023,dago_stabilizing_2024}.

The input-output relations for the atoms~\cref{eq:unitary_broadcast_io,eq:target_squeez,eq:two_qnd_unitary} show that the strength of the effective nonlinearity for the atoms is defined by the stiffness of the source nonlinearity $\gamma$ and the linear QND gain $\cG$.
On the one hand, this allows expecting a significant nonlinearity stiffness for the atomic system $\gamma \cG^3$ even given a weak mechanical nonlinearity $\gamma$.
In an experiment, the gain $\cG$ can be increased by increasing the interaction rates and/or duration of the local QND interactions with the optical mediator (atom-optical and opto-mechanical; the technical details of the implementation of the atom-mechanical QND gate are in~\cite{manukhova_pulsed_2020}).
Each of these ways has disadvantages along with increasing~$\cG$.
Larger duration $\tau$ causes a stronger influence of the thermal heating on the mechanical oscillator as the heating admixes additive noise proportional to~$\heating \tau$~($\heating$ is the heating rate).
Increasing either of the local QND coupling rates (atom-optical $g\s{a}$ and opto-mechanical $g\s m$), unfortunately, leads to stronger distortion of the temporal profile of the mediating pulse, which makes it impossible to mode-match it to both atomic and optomechanical cavities simultaneously.
This is equivalent to an increase in the mediator loss.
Increasing the gains~$\cG$ by any means is, therefore, associated with extra loss or noise, so a wise optimization of the control parameters is required, which we perform.
Given a fixed total gain, it appears to be more advantageous to use stronger optomechanical coupling $g\s m$ and weaker opto-atomic one $g\s a$ than to have both quantities equal, which is proven by numerical simulations.
The reason for this is in the inner workings of the hybrid QND gate (see~\cite{manukhova_pulsed_2020} for details).

The key elements of our broadcasting protocol are a linear QND gate connecting the source and the target system, and a nonlinearity of the source system.
As an example, we showed that with the resources that are within reach of current atom-optomechanical systems (e.g.,~\cite{thomas_entanglement_2020,karg_lightmediated_2020}), a cubic nonlinearity ($\propto \exp[ - \ii \xt^3 ]$) can be broadcast from the mechanics to the atoms.
The cubic nonlinearity considered here has the advantage of being non-demolition: it conserves the position quadrature of the source.
Such a nonlinearity can be broadcast to the atomic system in an advantageous way that enhances the effective strength of the nonlinearity using the gain of the linear QND coupling between the source and the target.
Similarly, other nondemolition nonlinearities of mechanical motion can be broadcast to atomic systems or other linearized bosonic systems that can be coupled via a linear QND interaction to the source mechanical mode.
Note that the property of being non-demolition is peculiar to the nonlinearities that originate from nonlinear potentials of the mechanical motion.
The non-demolition character is not necessary for an arbitrary nonlinearity to be broadcast.
In principle, the state-transfer stages of our protocol (the first two and the last two linear QND atom-mechanical interactions) can be arranged in a way that implements an approximate state swap between the systems.
This way, the initial state of the target is first transferred to the source, then processed in a nonlinear manner, and then transferred back to the target.
The state swap interaction is readily available for certain atomic (see Ref.~\cite{hammerer_quantum_2010} for a review) and optomechanical~\cite{chan_laser_2011,teufel_sideband_2011} systems.
For these systems, broadcasting using the state swap interactions right away may be optimal.
However, for other systems without the swap interactions, the proposed broadcasting strategy based on linear QND interactions is a viable solution.
The state-swap-based broadcasting, however, does not allow amplifying the nonlinearity in contrast to our present proposal.
Additionally, linear QND (sum) gates between different systems~\cite{manukhova_pulsed_2020} have additional universality and versatility for general quantum processing~\cite{gottesman_encoding_2001,braunstein_quantum_2005}, compared with the swap gates.
A proper comparison of the two strategies requires a full theoretical analysis of both schemes, including the imperfections, and lies beyond the scope of our manuscript.

Linearized bosonic description of atomic spin ensembles allows highly efficient generation, processing, and readout of their Gaussian quantum states~\cite{hammerer_quantum_2010}.
However, it is well known that applications of the Gaussian dynamics are fundamentally limited.
Such dynamics cannot enable universal quantum computation, fault tolerance, or full quantum advantage in metrology.
The inclusion of nonlinear quantum gates into the toolbox of atomic systems represents a qualitative extension of the operational toolbox to overcome these limitations.
Even weak nonlinearities promote the atomic systems beyond the Gaussian regime, allowing access to universal gate sets, enhanced sensing, and preparation of quantum non-Gaussian states, inaccessible to linearized Gaussian dynamics alone~\cite{lloyd_quantum_1999,braunstein_quantum_2005}.
Therefore, broadcasting the nonlinearity is essential for unlocking the full quantum potential of linearized spin ensembles considered as quantum memories in all the pillars of quantum technology. 

\section{Methods} \label{sec:methods}

\subsection{Input-output relations for the unitary gates}  \label{sec:input_output_relations_for_the_qnd_gates}

In this section we write down the input-output relations for the basic building gates of our protocol, and derive the input-output relations of~\cref{eq:4qnd_gate,eq:4qnd_gateb}.
This allows us to unambiguously specify the exact transformations that are implemented in the system.

The first type of the Gaussian gates is described by the Hamiltonian $H\s{qy} \propto \xs \pt$.
Such a gate preserves the non-demolition quadratures $\xs$ and $\pt$ intact, but induces displacements in the conjugate quadratures:
\begin{align}
  \xt & \mapsto \xt + g \xs,
  &
  \ps & \mapsto \ps - g \pt.
\end{align}
This is derived from the Heisenberg equations:
\begin{align}
  \diff{ \ps }{ t } & \propto \frac \ii 2 \comm{ \xs \pt }{ \ps } = \frac \ii 2 \pt \comm{ \xs }{ \ps } = - \pt,
  &
  \diff{\xt}{ t} & \propto \frac \ii 2  \comm{ \xs \pt }{ \xt } = \xs.
\end{align}
The second type of the Gaussian gates, described by the Hamiltonian $H\s{px} \propto \ps \xt$, does not change the non-demolition variables $\ps$ and $\xt$, but transforms the conjugate quadratures similarly:
\begin{align}
  \pt & \mapsto \pt - g \xs,
  &
  \xs & \mapsto \xs + g \xt.
\end{align}

These transformations can be neatly summarized in the vector form.
For convenience, we gather the quadrature operators of the bipartite source-target system into a vector $\mvec r = (\xt, \pt, \xs, \ps)^\intercal$.
Action of the Gaussian gates $\UU_{qy } (g) = \exp[ - \ii g \xs \pt / 2 ]$ and $\UU_{px } (g) = \exp[ - \ii g \ps \xt / 2]$ can be described by multiplying this vector by the corresponding matrix $A_{qy }(g)$ or $A_{px }(g)$:
\begin{align}
  \UU_{qy }(g):  \mvec r & \mapsto A_{qy}(g) \mvec r,
  &
  \UU_{px }(g):  \mvec r & \mapsto A_{px}(g) \mvec r.
\end{align}
The matrices read
\begin{align}
  A_{qy } (g) & =
  \begin{pmatrix}
    1 & 0   & g & 0 \\
    0 & 1   & 0 & 0 \\
    0 & 0   & 1 & 0 \\
    0 & - g & 0 & 1
  \end{pmatrix},
  &
  A_{px } (g) & =
  \begin{pmatrix}
    1 & 0 & 0 & 0   \\
    0 & 1 & 0 & - g \\
    g & 0 & 1 & 0   \\
    0 & 0 & 0 & 1
  \end{pmatrix}.
\end{align}
The nonlinear transformation $\exp[ - \ii \gamma V(\xs)/2]$, applied to the source system, changes the quadratures of the entire system $\mvec r$ as follows:
\begin{equation}
  \mvec r = (\xt, \pt, \xs , \ps )^\intercal
  \mapsto
  \mathcal N [\mvec r ]
  = (\xt, \pt,  \xs , \ps - \gamma V' (\xs) )^\intercal.
\end{equation}

Having introduced these notations, we can write the input-output relations for the two systems in the form
\begin{equation}
  \mvec r\s{i} \mapsto
  \mvec r\s{f} = A_{qy }(g_4) A_{px } (g_3)
  \mathcal N [
  A_{px } (g_2) A_{qy } (g_1) \mvec r\s{i}
  ].
\end{equation}
Here, according to the conventions of~\cref{fig:fig0-pdf}, $\mvec r\s{i}$ contains the quadratures before the protocol, and $\mvec r\s{f}$ contains the quadratures after the protocol.

\subsection{Nonlinear variance as an indicator of nonlinear correlations}  \label{sec:nonlinear_variance_as_a_figure_of_merit}

Application of nonlinear gates to bosonic systems produces non-linear correlations between their quadratures.
In general case, a gate defined by nonlinear function of the position $V(\xs)$, transforms the quadratures ($\xs, \ps$) as
\begin{align}
  \ee^{ - \ii \gamma V(\xs) /2 }
  \xs
  \ee^{ \ii \gamma V(\xs) /2 }
  & = \xs,
  &
  \ee^{ - \ii \gamma V(\xs) /2 }
  \ps
  \ee^{ \ii \gamma V(\xs) /2 }
  & = \ps - \gamma V' (\xs).
\end{align}
Therefore, the operator of momentum $\ps$ becomes correlated with the function of position $V'(\xs)$.
Consequently, the variance of a nonlinear combination of the quadratures $\ps + \lambda V'(\xs)$, where $\lambda$ is a real-valued parameter, becomes suppressed for certain values of $\lambda$.

For concreteness, let us illustrate this using the example of a cubic nonlinearity $V(x) = x^3/3$ and a mode with quadratures ($\xs, \ps$).
The corresponding nonlinear variance for this nonlinearity reads
\begin{multline}
  \label{eq:nlv_cubic_arbitrary}
  \nlv_3 (\lambda) = \Var( \ps + \lambda \xs^2 )
  = \ev{ \left( \ps + \lambda \xs^2 \right)^2 } - \ev{ \ps + \lambda \xs^2 }^2
  \\
= \Var( \ps ) + \lambda^2 \Var ( \xs^2 )
  + 2 \lambda \left( \frac 12 \ev{ \acomm{ \ps }{ \xs ^2 } } - \ev{ \ps } \ev{ \xs^2} \right).
\end{multline}
All the expectations are meant to be computed over the state being investigated.
By construction, the NLV is a quadratic parabola in $\lambda$.
Moreover, being a variance of a Hermitian operator, it is non-negative.

To illustrate how the properties of this parabola are related to the non-linear properties of the quantum state, it is instructive to compute the NLV for a few particular quantum states.
One instructive example is a squeezed thermal state, that is a thermal state to which a squeeze operator is applied.
The result is a Gaussian state with zero mean values of the quadratures, and a diagonal covariance matrix $( 2 \bar n + 1 ) \diag ( s, s ^{-1})$.
Here $\bar n$ is the mean occupation, $s$ is the squeezing magnitude.
For simplicity, we do not consider other possible squeezing phases.
Substituting the moments of the Gaussian state into~\cref{eq:nlv_cubic_arbitrary}, we obtain for the NLV of a squeezed thermal state (and the ground state as a limit of a non-squeezed thermal state with zero mean occupation)
\begin{equation}
  \nlv_3\up{ST} (\lambda) = \frac{ 2 \bar n + 1 }{ s } + 2
  \left( \lambda s ( 2 \bar n + 1 )\right)^2
  \qquad
  \underset{s=1,\bar n=0}{\longrightarrow}
  1 + 2 \lambda^2.
\end{equation}
We used that in a zero-mean Gaussian state, $\ev{\xs^4} = 3 \ev{\xs^2}^2$, and therefore $\Var( \xs^2 ) = \ev{ \xs^4 } - \ev{ \xs^2 }^2 = 2 \ev{ \xs^2 }^2$.

Therefore, for squeezed thermal states, the nonlinear variance is always a parabola in $\lambda$ with the minimum at $\lambda = 0$.
The value of the minimum decreases with decrease of the occupation $\bar n$ and also decreases with increase of the squeezing $s$.
The steepness of the parabola increases with both $s$ and $\bar n$.

Now, let us consider an application of a cubic nonlinearity to an arbitrary quantum state $\rho\s{i}$:
\begin{equation}
  \rho\s{f} = \ee^{ - \ii \gamma \xs^3 / 6 } \rho\s{i}\; \ee^{ \ii \gamma \xs^3 / 6 }.
\end{equation}
In the Heisenberg picture, the nonlinearity maps the quadratures as follows:
\begin{align}
  \xs \mapsto & \xs,
  &
  \ps \mapsto & \ps - \gamma \xs^2.
\end{align}
Using this in the definition of the NLV, one obtains
\begin{equation}
  \nlv_3^{ \rho\s{f}} (\lambda)
  = \Var( \ps + \lambda \xs^2 )_{\rho\s{f}}
  = \Var( ( \ps - \gamma\xs^2 ) + \lambda \xs^2 )_{\rho\s{i}}
  = \Var( \ps + ( \lambda - \gamma ) \xs^2 )
  = \nlv_3^{\rho\s{i}} (\lambda - \gamma).
\end{equation}
That is, in the NLV, application of the cubic nonlinearity simply shifts the NLV parabola along the $\lambda$-axis.
The displacement is determined by the strength $\gamma$ of the applied nonlinearity.
Therefore, the value of $\lambda$, at which the parabola attains the minimum value, indicates the effective strength of the cubic gate used to obtain the quantum state.

For the detection of quantum states produced by the cubic nonlinearity, the NLV appears to be an indispensable tool~\cite{moore_hierarchy_2022}.
It can be shown that for classical states
(such states $\hat \rho$ that can be represented as
\[
  \hat \rho = \int P(\alpha) \dyad{\alpha}{\alpha} \dd ^2 \alpha
\]
with Glauber-Sudarshan $P$-function being a regular probability distribution),
$\nlv (\lambda) \geq \nlv \s{NC} (\lambda)$ holds for arbitrary lambda.
Similarly, for all Gaussian quantum states (arbitrary convex mixtures of displaced squeezed vacua $\hat{ \mathcal D} \hat{ \mathcal S } \ket 0 $), $\nlv (\lambda) \geq \nlv\s{NG} (\lambda)$ for each lambda.
Correspondingly, if for any $\lambda$ the quantum state of the target system exhibits the value $\nlv (\lambda)$ below the value of the respective threshold, this state is non-classical and/or non-Gaussian.

A similarity to the detection of the Gaussian squeezing can be drawn.
A state of the mode with quadratures $(\xs, \ps)$ is said to be squeezed when
\begin{equation}
  \Var ( \xs \cos \theta + \ps \sin \theta ) < \sigma\s{vac},
\end{equation}
for at least one value of the phase $\theta$.
Here $\sigma\s{vac}$ is the shot-noise variance.
This inequality can be reexpressed as:
\begin{equation}
  \Var ( \xs \cos \theta + \ps \sin \theta ) = \sin^2\theta \Var ( \ps + \xs \tan \theta ) = \frac{ 1 }{ 1 + \lambda^2 } \Var( \ps + \lambda \xs ) < \sigma\s{vac}.
\end{equation}
Which yields a threshold $\sigma\s{S} (\lambda) = 1 + \lambda^2$ similar to~\cref{eq:nonlin_var_thrs} to discriminate squeezed and non-squeezed states.

Suppression of fluctuations in the nonlinear combination of canonical quadratures has practical applications in e.g. measurement-based quantum computing~\cite{marek_general_2018} where it serves as a resource.
The nonlinear squeezing appears to be a more appropriate indication of reaching the non-classicality or non-Gaussianity than quantum state fidelity, as the latter requires definition of the target state.

\subsection{Optimization of the envelopes of the nonlinear variance}  \label{sec:Variance_minimization}

By its construction, for each state $\hat \rho$, the NLV~$\nlv (\lambda)$ is a quadratic function of its argument.
The value $\lambda_0$ corresponding to the minimal value of $\nlv$, and the value $\nlv (\lambda_0)$ is determined by the state $\hat \rho$.
In our case, these properties are defined by the configuration of the system, i.e.\ system properties such as available nonlinearity $\gamma$, cavities' linewidths $\kappa_{\sbullet}$, local interaction rates $g_{\sbullet}$, damping $\zeta_{\sbullet}$ and heating $\heating$ rates, initial occupation $\bar n$ and squeezing $S\s m$ of the mechanics, and the mediator loss $\eta$, and the control parameters including the local QND coupling rates $g_{\sbullet}$, mediator pulse duration $\tau$, and the mediator squeezing $S$.
The nonlinearity is included in the properties of the system rather than the control parameters, as we consider it to be difficult to change.
This is in contrast to the control parameters which are assumed to be easy to change on-demand.

Varying the control parameters for each configuration of the system produces different parabolas $\nlv(\lambda)$.
We aim to determine the envelope of the minimum attainable values of the parabolas as a function of $\lambda$.
To this end, for each configuration of the system (each set $\{\gamma, \kappa\s{a,m}, \heating, n, S\s m, \eta\}$), for each value of $\lambda$, we minimize the resulting value of the NLV over the control parameters.
The result of this optimization is the envelope of the NLV parabolas:
\begin{equation}
    \nlv\up{ENV} (\lambda;  \gamma, \kappa\s{a,m}, \heating, n, S\s m, \eta)
    =
    \min_{g\s{a,m}, S, \tau} \nlv ( \lambda;  \gamma, \kappa\s{a,m}, \heating, n, S\s m, \eta, g\s{a,m}, S , \tau).
\end{equation}

In general, the parameter space is very large, and hence the optimization is very difficult.
Therefore, we enforce certain constraints on the control parameters to reduce the parameter space.

We start with the full protocol depicted in~\cref{fig:fig0-pdf}~(a).
First, we consider the \emph{gain-symmetric} regime of the QND gates.
In this regime, the local coupling rates of the atom-optical and optomechanical interactions are related in a way to establish equal gains $\cG\s a = \cG\s m = \cG$ in the QND transformations~\cref{eq:real_qnd,eq:real_qnd_a}.
This way, each QND transformation is parametrized by a single value of the gain $\cG$.
Furthermore, we maintain the relationship between the individual $\cG_{\sbullet}$ such that it corresponds to the regime of the nonlinearity broadcasting (see~\cref{eq:unitary_broadcast_io}).
The resulting envelopes are shown with dashed lines in~\cref{fig:2&4qnd_nlv_real}~(a).

Second, we consider the \emph{gain-asymmetric} regime, where each atom-mechanical QND gate is parametrized by two independent gains $\cG\s a \neq \cG\s m$.
The relationship between the gains of different QND gates corresponding to the broadcasting regime is still maintained.
The result of this optimization is shown with thick solid lines in~\cref{fig:2&4qnd_nlv_real}~(a).
We can see that thanks to the smaller constraints, the gain-asymmetric regime is superior compared the gain-symmetric one.

Finally, we optimize over all the gains $\cG_{\sbullet}$ assuming them independent.
The result of this optimization essentially recovers the regime of the generation of nonlinear squeezing.
It is shown with thin solid lines in~\cref{fig:2&4qnd_nlv_real}~(a).
We have to note that in the lossless case $\eta = 1$, the QND gains $\cG\s{a,m}$ are equal right away.
Therefore, in this case, gain-symmetric and gain-asymmetric regime are exactly the same.

The simplified protocol is treated in a very similar manner.
We first perform its optimization assuming the gain-symmetric regime of atom-mechanical QND gates, and then gain-asymmetric.
The simplified protocol does not have separate regimes of broadcasting/ nonlinear squeezing generation.
Instead, we directly optimize the control parameters to minimize the NLV envelopes, with the results presented on~\cref{fig:2&4qnd_nlv_real}~(b).

\subsection{Numerical simulation of the quantum dynamics}  \label{sec:numerical_lindblad}

In order to simulate the Wigner functions of quantum states at different stages of the protocol, we use the open-source software QuTiP~\cite{johansson_qutip_2012,johansson_qutip_2013,lambert_qutip_2024}.

The quantum state of the atom-mechanical system is represented by its density matrix $\rho$ in the Fock basis.
The atom-mechanical QND gates are described by a master equation in Lindblad form
\begin{align}
  \diff{\rho}{t} = - \frac{ \ii }{ \hbar } \comm{ H }{ \rho } +
  \zeta\s a \mathcal D [ a\s a ] \rho + \zeta\s m ( n\s{th} + 1 ) \mathcal D [ a\s m ] \rho + \zeta \s m n\s{th} \mathcal D [a \s m^\dag] \rho,
\end{align}
where $a\s{a,m}$ are the annihilation operators, $\zeta\s{a,m}$ denote the decay rates of atoms and mechanics, and $n\s{th}$ is the mean occupation of the mechanical environment, so that the heating rate of the mechanics is given by $\heating \equiv \zeta\s{m} n\s{th}$
The superoperators $\mathcal D$ are given by
\begin{equation}
  \mathcal D [c] \rho \equiv c \rho c^\dag - \frac 12 \mleft[ c^\dag c \rho + \rho c^\dag c \mright].
\end{equation}

The intermediate step of the nonlinear evolution can be considered a unitary transformation thanks to its short duration.
This allows us to reduce the computation of the nonlinear transformation to the application of a unitary map
\begin{equation}
  \rho \mapsto e^{ - \ii \gamma X\s m^3 / 6 } \rho e^{ \ii \gamma X\s m ^3 / 6 }.
\end{equation}

\section*{Data Availability} 

Data sharing not applicable~--- the article describes entirely theoretical research and numerical simulations.

\section*{Acknowledgements} 

\sloppy

The authors are grateful to Darren W. Moore for helpful discussions.
We acknowledge the project 23-06224S of the Czech Science Foundation and project CZ.02.01.01/00/22\_008/0004649 of the MEYS of the Czech Republic supported by the EU funding.
A.R. and R.F. have been also supported by the project No. LUC25006 of MEYS Czech Republic.

\section*{Competing Interests} 

The authors declare no competing interests.

\section*{Author Contributions} 

R.F. developed theoretical idea and supervised the project. A.D.M. performed calculations and made the figures  with input from A.A.R. and R.F.
A.A.R. and A.D.M prepared the manuscript with inputs from R.F.
All authors jointly contributed to analysis and discussion of the manuscript.

\bibliographystyle{unsrt}
\bibliography{BiblioNonLin,refs_andrey}

\begin{thebibliography}{100}

\bibitem{lloyd_quantum_1999}
Seth Lloyd and Samuel~L. Braunstein.
\newblock Quantum {{Computation}} over {{Continuous Variables}}.
\newblock {\em Physical Review Letters}, 82(8):1784--1787, February 1999.

\bibitem{braunstein_quantum_2005}
Samuel Braunstein and Peter {van Loock}.
\newblock Quantum information with continuous variables.
\newblock {\em Reviews of Modern Physics}, 77(2):513--577, June 2005.

\bibitem{kendon_quantum_2010}
Vivien~M. Kendon, Kae Nemoto, and William~J. Munro.
\newblock Quantum analogue computing.
\newblock {\em Philosophical Transactions of the Royal Society A: Mathematical,
  Physical and Engineering Sciences}, 368(1924):3609--3620, August 2010.

\bibitem{georgescu_quantum_2014}
I.~M. Georgescu, S.~Ashhab, and Franco Nori.
\newblock Quantum simulation.
\newblock {\em Reviews of Modern Physics}, 86(1):153--185, March 2014.

\bibitem{cerf_quantum_2007}
Nicolas~J. Cerf, Gerd Leuchs, and Eugene~S. Polzik, editors.
\newblock {\em Quantum {{Information With Continuous Variables Of Atoms And
  Light}}}.
\newblock Imperial College Press, London : Hackensack, N.J, February 2007.

\bibitem{polzik_quantum_2008}
Eugene~S. Polzik and Jarom{\'i}r Fiur{\'a}{\v s}ek.
\newblock Quantum {{Interface Between Light}} and {{Atomic Ensembles}}.
\newblock In {\em Lectures on {{Quantum Information}}}, pages 515--535.
  Wiley-Blackwell, 2008.

\bibitem{treutlein_hybrid_2014}
Philipp Treutlein, Claudiu Genes, Klemens Hammerer, Martino Poggio, and Peter
  Rabl.
\newblock Hybrid {{Mechanical Systems}}.
\newblock In Markus Aspelmeyer, Tobias~J. Kippenberg, and Florian Marquardt,
  editors, {\em Cavity {{Optomechanics}}: {{Nano-}} and {{Micromechanical
  Resonators Interacting}} with {{Light}}}, pages 327--351. Springer Berlin
  Heidelberg, Berlin, Heidelberg, 2014.

\bibitem{clerk_hybrid_2020}
A.~A. Clerk, K.~W. Lehnert, P.~Bertet, J.~R. Petta, and Y.~Nakamura.
\newblock Hybrid quantum systems with circuit quantum electrodynamics.
\newblock {\em Nature Physics}, pages 257--267, March 2020.

\bibitem{chu_perspective_2020}
Yiwen Chu and Simon Gr{\"o}blacher.
\newblock A perspective on hybrid quantum opto- and electromechanical systems.
\newblock {\em Applied Physics Letters}, 117(15):150503, October 2020.

\bibitem{moller_quantum_2017}
Christoffer~B. M{\o}ller, Rodrigo~A. Thomas, Georgios Vasilakis, Emil Zeuthen,
  Yeghishe Tsaturyan, Mikhail Balabas, Kasper Jensen, Albert Schliesser,
  Klemens Hammerer, and Eugene~S. Polzik.
\newblock Quantum back-action-evading measurement of motion in a negative mass
  reference frame.
\newblock {\em Nature}, 547(7662):191--195, July 2017.

\bibitem{karg_remote_2019}
Thomas~M. Karg, Baptiste Gouraud, Philipp Treutlein, and Klemens Hammerer.
\newblock Remote {{Hamiltonian}} interactions mediated by light.
\newblock {\em Physical Review A}, 99(6):063829, June 2019.

\bibitem{karg_lightmediated_2020}
Thomas~M. Karg, Baptiste Gouraud, Chun~Tat Ngai, Gian-Luca Schmid, Klemens
  Hammerer, and Philipp Treutlein.
\newblock Light-mediated strong coupling between a mechanical oscillator and
  atomic spins 1 meter apart.
\newblock {\em Science}, May 2020.

\bibitem{thomas_entanglement_2020}
Rodrigo~A. Thomas, Micha{\l} Parniak, Christoffer {\O}stfeldt, Christoffer~B.
  M{\o}ller, Christian B{\ae}rentsen, Yeghishe Tsaturyan, Albert Schliesser,
  J{\"u}rgen Appel, Emil Zeuthen, and Eugene~S. Polzik.
\newblock Entanglement between distant macroscopic mechanical and spin systems.
\newblock {\em Nature Physics}, pages 1--6, September 2020.

\bibitem{aspelmeyer_cavity_2014}
Markus Aspelmeyer, Tobias~J. Kippenberg, and Florian Marquardt.
\newblock Cavity optomechanics.
\newblock {\em Reviews of Modern Physics}, 86(4):1391--1452, December 2014.

\bibitem{teufel_sideband_2011}
J.~D. Teufel, T.~Donner, Dale Li, J.~W. Harlow, M.~S. Allman, K.~Cicak, A.~J.
  Sirois, J.~D. Whittaker, K.~W. Lehnert, and R.~W. Simmonds.
\newblock Sideband cooling of micromechanical motion to the quantum ground
  state.
\newblock {\em Nature}, 475(7356):359--363, July 2011.

\bibitem{chan_laser_2011}
Jasper Chan, T.~P.~Mayer Alegre, Amir~H. {Safavi-Naeini}, Jeff~T. Hill, Alex
  Krause, Simon Groeblacher, Markus Aspelmeyer, and Oskar Painter.
\newblock Laser cooling of a nanomechanical oscillator into its quantum ground
  state.
\newblock {\em Nature}, 478(7367):89--92, October 2011.

\bibitem{delic_cooling_2020}
Uro{\v s} Deli{\'c}, Manuel Reisenbauer, Kahan Dare, David Grass, Vladan
  Vuleti{\'c}, Nikolai Kiesel, and Markus Aspelmeyer.
\newblock Cooling of a levitated nanoparticle to the motional quantum ground
  state.
\newblock {\em Science}, 367(6480):892--895, February 2020.

\bibitem{piotrowski_simultaneous_2023}
Johannes Piotrowski, Dominik Windey, Jayadev Vijayan, Carlos
  {Gonzalez-Ballestero}, Andr{\'e}s {de los R{\'i}os Sommer}, Nadine Meyer,
  Romain Quidant, Oriol {Romero-Isart}, Ren{\'e} Reimann, and Lukas Novotny.
\newblock Simultaneous ground-state cooling of two mechanical modes of a
  levitated nanoparticle.
\newblock {\em Nature Physics}, 19(7):1009--1013, July 2023.

\bibitem{wollman_quantum_2015}
E.~E. Wollman, C.~U. Lei, A.~J. Weinstein, J.~Suh, A.~Kronwald, F.~Marquardt,
  A.~A. Clerk, and K.~C. Schwab.
\newblock Quantum squeezing of motion in a mechanical resonator.
\newblock {\em Science}, 349(6251):952--955, August 2015.

\bibitem{pirkkalainen_squeezing_2015}
J.-M. Pirkkalainen, E.~Damsk{\"a}gg, M.~Brandt, F.~Massel, and M.~A.
  Sillanp{\"a}{\"a}.
\newblock Squeezing of {{Quantum Noise}} of {{Motion}} in a {{Micromechanical
  Resonator}}.
\newblock {\em Physical Review Letters}, 115(24):243601, December 2015.

\bibitem{safavi-naeini_squeezed_2013}
Amir~H. {Safavi-Naeini}, Simon Gr{\"o}blacher, Jeff~T. Hill, Jasper Chan,
  Markus Aspelmeyer, and Oskar Painter.
\newblock Squeezed light from a silicon micromechanical resonator.
\newblock {\em Nature}, 500(7461):185--189, August 2013.

\bibitem{ockeloen-korppi_noiseless_2017}
C.~F. {Ockeloen-Korppi}, E.~Damsk{\"a}gg, J.-M. Pirkkalainen, T.~T.
  Heikkil{\"a}, F.~Massel, and M.~A. Sillanp{\"a}{\"a}.
\newblock Noiseless {{Quantum Measurement}} and {{Squeezing}} of {{Microwave
  Fields Utilizing Mechanical Vibrations}}.
\newblock {\em Physical Review Letters}, 118(10):103601, March 2017.

\bibitem{militaru_ponderomotive_2022}
Andrei Militaru, Massimiliano Rossi, Felix Tebbenjohanns, Oriol {Romero-Isart},
  Martin Frimmer, and Lukas Novotny.
\newblock Ponderomotive {{Squeezing}} of {{Light}} by a {{Levitated
  Nanoparticle}} in {{Free Space}}.
\newblock {\em Physical Review Letters}, 129(5):053602, July 2022.

\bibitem{palomaki_entangling_2013}
T.~A. Palomaki, J.~D. Teufel, R.~W. Simmonds, and K.~W. Lehnert.
\newblock Entangling {{Mechanical Motion}} with {{Microwave Fields}}.
\newblock {\em Science}, 342(6159):710--713, August 2013.

\bibitem{riedinger_nonclassical_2016}
Ralf Riedinger, Sungkun Hong, Richard~A. Norte, Joshua~A. Slater, Juying Shang,
  Alexander~G. Krause, Vikas Anant, Markus Aspelmeyer, and Simon
  Gr{\"o}blacher.
\newblock Non-classical correlations between single photons and phonons from a
  mechanical oscillator.
\newblock {\em Nature}, 530(7590):313--316, February 2016.

\bibitem{riedinger_remote_2018}
Ralf Riedinger, Andreas Wallucks, Igor Marinkovi{\'c}, Clemens L{\"o}schnauer,
  Markus Aspelmeyer, Sungkun Hong, and Simon Gr{\"o}blacher.
\newblock Remote quantum entanglement between two micromechanical oscillators.
\newblock {\em Nature}, 556(7702):473--477, April 2018.

\bibitem{ockeloen-korppi_stabilized_2018}
C.~F. {Ockeloen-Korppi}, E.~Damsk{\"a}gg, J.-M. Pirkkalainen, M.~Asjad, A.~A.
  Clerk, F.~Massel, M.~J. Woolley, and M.~A. Sillanp{\"a}{\"a}.
\newblock Stabilized entanglement of massive mechanical oscillators.
\newblock {\em Nature}, 556(7702):478--482, April 2018.

\bibitem{ranjit_zeptonewton_2016}
Gambhir Ranjit, Mark Cunningham, Kirsten Casey, and Andrew~A. Geraci.
\newblock Zeptonewton force sensing with nanospheres in an optical lattice.
\newblock {\em Physical Review A}, 93(5):053801, May 2016.

\bibitem{dominguez-medina_neutral_2018}
Sergio {Dominguez-Medina}, Shawn Fostner, Martial Defoort, Marc Sansa,
  Ann-Kathrin Stark, Mohammad~Abdul Halim, Emeline Vernhes, Marc Gely,
  Guillaume Jourdan, Thomas Alava, Pascale Boulanger, Christophe Masselon, and
  S{\'e}bastien Hentz.
\newblock Neutral mass spectrometry of virus capsids above 100 megadaltons with
  nanomechanical resonators.
\newblock {\em Science}, 362(6417):918--922, November 2018.

\bibitem{mason_continuous_2019}
David Mason, Junxin Chen, Massimiliano Rossi, Yeghishe Tsaturyan, and Albert
  Schliesser.
\newblock Continuous force and displacement measurement below the standard
  quantum limit.
\newblock {\em Nature Physics}, page~1, May 2019.

\bibitem{carney_mechanical_2021}
Daniel Carney, Gordan Krnjaic, David~C. Moore, Cindy~A. Regal, Gadi Afek, Sunil
  Bhave, Benjamin Brubaker, Thomas Corbitt, Jonathan Cripe, Nicole Crisosto,
  Andrew Geraci, Sohitri Ghosh, Jack G.~E. Harris, Anson Hook, Edward~W. Kolb,
  Jonathan Kunjummen, Rafael~F. Lang, Tongcang Li, Tongyan Lin, Zhen Liu,
  Joseph Lykken, Lorenzo Magrini, Jack Manley, Nobuyuki Matsumoto, Alissa
  Monte, Fernando Monteiro, Thomas Purdy, C.~Jess Riedel, Robinjeet Singh,
  Swati Singh, Kanupriya Sinha, Jacob~M. Taylor, Juehang Qin, Dalziel~J.
  Wilson, and Yue Zhao.
\newblock Mechanical {{Quantum Sensing}} in the {{Search}} for {{Dark Matter}}.
\newblock {\em Quantum Science and Technology}, 6(2):024002, April 2021.

\bibitem{carney_searches_2023}
Daniel Carney, Kyle~G. Leach, and David~C. Moore.
\newblock Searches for {{Massive Neutrinos}} with {{Mechanical Quantum
  Sensors}}.
\newblock {\em PRX Quantum}, 4(1):010315, February 2023.

\bibitem{heinze_darkgeo_2024}
Joscha Heinze, Alex Gill, Artemiy Dmitriev, Ji{\v r}{\'i} Smetana, Tianliang
  Yan, Vincent Boyer, Denis Martynov, Hartmut Grote, James Lough, Aldo Ejlli,
  and Guido M{\"u}ller.
\newblock {{DarkGEO}}: A large-scale laser-interferometric axion detector.
\newblock {\em New Journal of Physics}, 26(5):055002, May 2024.

\bibitem{baker_optomechanical_2024}
Christopher~G. Baker, Warwick~P. Bowen, Peter Cox, Matthew~J. Dolan, Maxim
  Goryachev, and Glen Harris.
\newblock Optomechanical dark matter instrument for direct detection.
\newblock {\em Physical Review D}, 110(4):043005, August 2024.

\bibitem{bagci_optical_2014}
T.~Bagci, A.~Simonsen, S.~Schmid, L.~G. Villanueva, E.~Zeuthen, J.~Appel, J.~M.
  Taylor, A.~S{\o}rensen, K.~Usami, A.~Schliesser, and E.~S. Polzik.
\newblock Optical detection of radio waves through a nanomechanical transducer.
\newblock {\em Nature}, 507(7490):81--85, March 2014.

\bibitem{higginbotham_harnessing_2018}
A.~P. Higginbotham, P.~S. Burns, M.~D. Urmey, R.~W. Peterson, N.~S. Kampel,
  B.~M. Brubaker, G.~Smith, K.~W. Lehnert, and C.~A. Regal.
\newblock Harnessing electro-optic correlations in an efficient mechanical
  converter.
\newblock {\em Nature Physics}, 14(10):1038--1042, October 2018.

\bibitem{gonzalez-ballestero_levitodynamics_2021}
C.~{Gonzalez-Ballestero}, M.~Aspelmeyer, L.~Novotny, R.~Quidant, and
  O.~{Romero-Isart}.
\newblock Levitodynamics: {{Levitation}} and control of microscopic objects in
  vacuum.
\newblock {\em Science}, 374(6564):eabg3027, October 2021.

\bibitem{siler_diffusing_2018}
Martin {\v S}iler, Luca Ornigotti, Oto Brzobohat{\'y}, Petr J{\'a}kl, Artem
  Ryabov, Viktor Holubec, Pavel Zem{\'a}nek, and Radim Filip.
\newblock Diffusing up the {{Hill}}: {{Dynamics}} and {{Equipartition}} in
  {{Highly Unstable Systems}}.
\newblock {\em Physical Review Letters}, 121(23):230601, December 2018.

\bibitem{gutierrezlatorre_superconducting_2023}
Mart{\'i} Gutierrez~Latorre, Gerard Higgins, Achintya Paradkar, Thilo Bauch,
  and Witlef Wieczorek.
\newblock Superconducting {{Microsphere Magnetically Levitated}} in an
  {{Anharmonic Potential}} with {{Integrated Magnetic Readout}}.
\newblock {\em Physical Review Applied}, 19(5):054047, May 2023.

\bibitem{rakhubovsky_stroboscopic_2021}
Andrey~A. Rakhubovsky and Radim Filip.
\newblock Stroboscopic high-order nonlinearity for quantum optomechanics.
\newblock {\em npj Quantum Information}, 7(1):120, July 2021.

\bibitem{neumeier_fast_2024}
Lukas Neumeier, Mario~A. Ciampini, Oriol {Romero-Isart}, Markus Aspelmeyer, and
  Nikolai Kiesel.
\newblock Fast quantum interference of a nanoparticle via optical potential
  control.
\newblock {\em Proceedings of the National Academy of Sciences},
  121(4):e2306953121, January 2024.

\bibitem{roda-llordes_macroscopic_2024}
M.~{Roda-Llordes}, A.~{Riera-Campeny}, D.~Candoli, P.~T. Grochowski, and
  O.~{Romero-Isart}.
\newblock Macroscopic {{Quantum Superpositions}} via {{Dynamics}} in a {{Wide
  Double-Well Potential}}.
\newblock {\em Physical Review Letters}, 132(2):023601, January 2024.

\bibitem{casulleras_optimization_2024}
Silvia Casulleras, Piotr~T. Grochowski, and Oriol {Romero-Isart}.
\newblock Optimization of static potentials for large delocalization and
  non-{{Gaussian}} quantum dynamics of levitated nanoparticles under
  decoherence.
\newblock {\em Physical Review A}, 110(3):033511, September 2024.

\bibitem{peano_nonlinear_2006}
V.~Peano and M.~Thorwart.
\newblock Nonlinear response of a driven vibrating nanobeam in the quantum
  regime.
\newblock {\em New Journal of Physics}, 8(2):21, February 2006.

\bibitem{leijssen_nonlinear_2017}
Rick Leijssen, Giada~R. La~Gala, Lars Freisem, Juha~T. Muhonen, and Ewold
  Verhagen.
\newblock Nonlinear cavity optomechanics with nanomechanical thermal
  fluctuations.
\newblock {\em Nature Communications}, 8(1):ncomms16024, July 2017.

\bibitem{guerra_electrostatically_2008}
Diego~N. Guerra, Matthias Imboden, and Pritiraj Mohanty.
\newblock Electrostatically actuated silicon-based nanomechanical switch at
  room temperature.
\newblock {\em Applied Physics Letters}, 93(3):033515, July 2008.

\bibitem{jacobs_engineering_2009}
Kurt Jacobs and Andrew~J. Landahl.
\newblock Engineering {{Giant Nonlinearities}} in {{Quantum Nanosystems}}.
\newblock {\em Physical Review Letters}, 103(6):067201, August 2009.

\bibitem{matheny_nonlinear_2013}
M.~H. Matheny, L.~G. Villanueva, R.~B. Karabalin, J.~E. Sader, and M.~L.
  Roukes.
\newblock Nonlinear {{Mode-Coupling}} in {{Nanomechanical Systems}}.
\newblock {\em Nano Letters}, 13(4):1622--1626, April 2013.

\bibitem{rips_nonlinear_2014}
S.~Rips, I.~{Wilson-Rae}, and M.~J. Hartmann.
\newblock Nonlinear nanomechanical resonators for quantum optoelectromechanics.
\newblock {\em Physical Review A}, 89(1):013854, January 2014.

\bibitem{ochs_amplification_2021}
J.~S. Ochs, M.~Seitner, M.~I. Dykman, and E.~M. Weig.
\newblock Amplification and spectral evidence of squeezing in the response of a
  strongly driven nanoresonator to a probe field.
\newblock {\em Physical Review A}, 103(1):013506, January 2021.

\bibitem{ochs_frequency_2022}
J.~S. Ochs, D.~K.~J. Bone{\ss}, G.~Rastelli, M.~Seitner, W.~Belzig, M.~I.
  Dykman, and E.~M. Weig.
\newblock Frequency {{Comb}} from a {{Single Driven Nonlinear Nanomechanical
  Mode}}.
\newblock {\em Physical Review X}, 12(arXiv:2207.04030):041019, November 2022.

\bibitem{rosiek_quadrature_2023}
Christian~A. Rosiek, Massimiliano Rossi, Albert Schliesser, and Anders~S.
  S{\o}rensen.
\newblock Quadrature squeezing enhances {{Wigner}} negativity in a mechanical
  {{Duffing}} oscillator, December 2023.

\bibitem{chu_cold_2002}
Steven Chu.
\newblock Cold atoms and quantum control.
\newblock {\em Nature}, 416:206, March 2002.

\bibitem{kitching_atomic_2011}
John Kitching, Svenja Knappe, and Elizabeth~A. Donley.
\newblock Atomic {{Sensors}} -- {{A Review}}.
\newblock {\em IEEE Sensors Journal}, 11(9):1749--1758, September 2011.

\bibitem{monroe_quantum_2002}
C.~Monroe.
\newblock Quantum information processing with atoms and photons.
\newblock {\em Nature}, 416(6877):238--246, March 2002.

\bibitem{lukin_colloquium_2003}
M.~D. Lukin.
\newblock Colloquium: {{Trapping}} and manipulating photon states in atomic
  ensembles.
\newblock {\em Reviews of Modern Physics}, 75(2):457--472, April 2003.

\bibitem{kimble_quantum_2008}
H.~J. Kimble.
\newblock The quantum internet.
\newblock {\em Nature}, 453(7198):1023--1030, June 2008.

\bibitem{sangouard_quantum_2011}
Nicolas Sangouard, Christoph Simon, Hugues {de Riedmatten}, and Nicolas Gisin.
\newblock Quantum repeaters based on atomic ensembles and linear optics.
\newblock {\em Reviews of Modern Physics}, 83(1):33--80, March 2011.

\bibitem{vanderwal_atomic_2003}
C.~H. {van der Wal}, M.~D. Eisaman, A.~Andr{\'e}, R.~L. Walsworth, D.~F.
  Phillips, A.~S. Zibrov, and M.~D. Lukin.
\newblock Atomic {{Memory}} for {{Correlated Photon States}}.
\newblock {\em Science}, 301(5630):196--200, July 2003.

\bibitem{julsgaard_experimental_2004}
Brian Julsgaard, Jacob Sherson, J.~Ignacio Cirac, Jarom{\'i}r Fiur{\'a}{\v
  s}ek, and Eugene~S. Polzik.
\newblock Experimental demonstration of quantum memory for light.
\newblock {\em Nature}, 432(7016):482--486, November 2004.

\bibitem{laurat_efficient_2006}
Julien Laurat, Hugues de~Riedmatten, Daniel Felinto, Chin-Wen Chou, Erik~W.
  Schomburg, and H.~Jeff Kimble.
\newblock Efficient retrieval of a single excitation stored in an atomic
  ensemble.
\newblock {\em Optics Express}, 14(15):6912--6918, July 2006.

\bibitem{simon_interfacing_2007}
Jonathan Simon, Haruka Tanji, James~K. Thompson, and Vladan Vuleti{\'c}.
\newblock Interfacing {{Collective Atomic Excitations}} and {{Single Photons}}.
\newblock {\em Physical Review Letters}, 98(18):183601, May 2007.

\bibitem{honda_storage_2008}
Kazuhito Honda, Daisuke Akamatsu, Manabu Arikawa, Yoshihiko Yokoi, Keiichirou
  Akiba, Satoshi Nagatsuka, Takahito Tanimura, Akira Furusawa, and Mikio
  Kozuma.
\newblock Storage and {{Retrieval}} of a {{Squeezed Vacuum}}.
\newblock {\em Physical Review Letters}, 100(9):093601, March 2008.

\bibitem{appel_quantum_2008}
Juergen Appel, Eden Figueroa, Dmitry Korystov, M.~Lobino, and A.~I. Lvovsky.
\newblock Quantum memory for squeezed light.
\newblock {\em Physical Review Letters}, 100(9):093602, March 2008.

\bibitem{choi_mapping_2008}
K.~S. Choi, H.~Deng, J.~Laurat, and H.~J. Kimble.
\newblock Mapping photonic entanglement into and out of a quantum memory.
\newblock {\em Nature}, 452(7183):67--71, March 2008.

\bibitem{yang_efficient_2016}
Sheng-Jun Yang, Xu-Jie Wang, Xiao-Hui Bao, and Jian-Wei Pan.
\newblock An efficient quantum light--matter interface with sub-second
  lifetime.
\newblock {\em Nature Photonics}, 10(6):381--384, June 2016.

\bibitem{saunders_cavityenhanced_2016}
D.~J. Saunders, J.~H.~D. Munns, T.~F.~M. Champion, C.~Qiu, K.~T. Kaczmarek,
  E.~Poem, P.~M. Ledingham, I.~A. Walmsley, and J.~Nunn.
\newblock Cavity-{{Enhanced Room-Temperature Broadband Raman Memory}}.
\newblock {\em Physical Review Letters}, 116(9):090501, March 2016.

\bibitem{wang_efficient_2019}
Yunfei Wang, Jianfeng Li, Shanchao Zhang, Keyu Su, Yiru Zhou, Kaiyu Liao,
  Shengwang Du, Hui Yan, and Shi-Liang Zhu.
\newblock Efficient quantum memory for single-photon polarization qubits.
\newblock {\em Nature Photonics}, 13(5):346--351, May 2019.

\bibitem{divincenzo_physical_2000}
David~P. DiVincenzo.
\newblock The {{Physical Implementation}} of {{Quantum Computation}}.
\newblock {\em Fortschritte der Physik}, 48(9-11):771--783, 2000.

\bibitem{anderson_observation_1995}
M.~H. Anderson, J.~R. Ensher, M.~R. Matthews, C.~E. Wieman, and E.~A. Cornell.
\newblock Observation of {{Bose-Einstein Condensation}} in a {{Dilute Atomic
  Vapor}}.
\newblock {\em Science}, 269(5221):198--201, July 1995.

\bibitem{davis_boseeinstein_1995}
K.~B. Davis, M.~O. Mewes, M.~R. Andrews, N.~J. {van Druten}, D.~S. Durfee,
  D.~M. Kurn, and W.~Ketterle.
\newblock Bose-{{Einstein Condensation}} in a {{Gas}} of {{Sodium Atoms}}.
\newblock {\em Physical Review Letters}, 75(22):3969--3973, November 1995.

\bibitem{pethick_bose_2008}
C.~J. Pethick and H.~Smith.
\newblock {\em Bose--{{Einstein Condensation}} in {{Dilute Gases}}}.
\newblock Cambridge University Press, Cambridge ; New York, 2nd edition
  edition, October 2008.

\bibitem{eriksson_universal_2024}
Axel~M. Eriksson, Th{\'e}o S{\'e}pulcre, Mikael Kervinen, Timo Hillmann, Marina
  Kudra, Simon Dupouy, Yong Lu, Maryam Khanahmadi, Jiaying Yang, Claudia
  {Castillo-Moreno}, Per Delsing, and Simone Gasparinetti.
\newblock Universal control of a bosonic mode via drive-activated native cubic
  interactions.
\newblock {\em Nature Communications}, 15(1):2512, March 2024.

\bibitem{manukhova_pulsed_2020}
A.~D. Manukhova, A.~A. Rakhubovsky, and R.~Filip.
\newblock Pulsed atom-mechanical quantum non-demolition gate.
\newblock {\em npj Quantum Information}, 6(1):4, January 2020.

\bibitem{shomroni_optical_2019}
Itay Shomroni, Liu Qiu, Daniel Malz, Andreas Nunnenkamp, and Tobias~J.
  Kippenberg.
\newblock Optical backaction-evading measurement of a mechanical oscillator.
\newblock {\em Nature Communications}, 10(1):2086, May 2019.

\bibitem{liu_quantum_2022}
Wang-Yan Liu, Li-Bao Fan, Ye-Xiong Zeng, Jin-Feng Huang, and Jie-Qiao Liao.
\newblock Quantum thermalization and thermal entanglement in the open quantum
  {{Rabi}} model.
\newblock {\em arXiv:2205.02676 [quant-ph]}, May 2022.

\bibitem{julsgaard_experimental_2001}
Brian Julsgaard, Alexander Kozhekin, and Eugene~S. Polzik.
\newblock Experimental long-lived entanglement of two macroscopic objects.
\newblock {\em Nature}, 413(6854):400--403, September 2001.

\bibitem{marek_general_2018}
Petr Marek, Radim Filip, Hisashi Ogawa, Atsushi Sakaguchi, Shuntaro Takeda,
  Jun-ichi Yoshikawa, and Akira Furusawa.
\newblock General implementation of arbitrary nonlinear quadrature phase gates.
\newblock {\em Physical Review A}, 97(2):022329, February 2018.

\bibitem{roda-llordes_numerical_2024}
M.~{Roda-Llordes}, D.~Candoli, P.~T. Grochowski, A.~{Riera-Campeny},
  T.~Agrenius, J.~J. {Garc{\'i}a-Ripoll}, C.~{Gonzalez-Ballestero}, and
  O.~{Romero-Isart}.
\newblock Numerical simulation of large-scale nonlinear open quantum mechanics.
\newblock {\em Physical Review Research}, 6(1):013262, March 2024.

\bibitem{budinger_alloptical_2024}
Niklas Budinger, Akira Furusawa, and Peter {van Loock}.
\newblock All-optical quantum computing using cubic phase gates.
\newblock {\em Physical Review Research}, 6(2):023332, June 2024.

\bibitem{moore_hierarchy_2022}
Darren~W. Moore and Radim Filip.
\newblock Hierarchy of quantum non-{{Gaussian}} conservative motion.
\newblock {\em Communications Physics}, 5(1):1--7, May 2022.

\bibitem{hammerer_quantum_2010}
Klemens Hammerer, Anders~S. S{\o}rensen, and Eugene~S. Polzik.
\newblock Quantum interface between light and atomic ensembles.
\newblock {\em Reviews of Modern Physics}, 82(2):1041--1093, April 2010.

\bibitem{siler_thermally_2017}
Martin {\v S}iler, Petr J{\'a}kl, Oto Brzobohat{\'y}, Artem Ryabov, Radim
  Filip, and Pavel Zem{\'a}nek.
\newblock Thermally induced micro-motion by inflection in optical potential.
\newblock {\em Scientific Reports}, 7(1):1697, May 2017.

\bibitem{moore_estimation_2019}
Darren~W. Moore, Andrey~A. Rakhubovsky, and Radim Filip.
\newblock Estimation of squeezing in a nonlinear quadrature of a mechanical
  oscillator.
\newblock {\em New Journal of Physics}, 21(11):113050, November 2019.

\bibitem{dania_ultrahigh_2024}
Lorenzo Dania, Dmitry~S. Bykov, Florian Goschin, Markus Teller, Abderrahmane
  Kassid, and Tracy~E. Northup.
\newblock Ultrahigh {{Quality Factor}} of a {{Levitated Nanomechanical
  Oscillator}}.
\newblock {\em Physical Review Letters}, 132(13):133602, March 2024.

\bibitem{braginsky_quantum_1980}
Vladimir~B. Braginsky, Yuri~I. Vorontsov, and Kip~S. Thorne.
\newblock Quantum {{Nondemolition Measurements}}.
\newblock {\em Science}, 209(4456):547--557, August 1980.

\bibitem{kala_cubic_2022}
Vojt{\v e}ch Kala, Radim Filip, and Petr Marek.
\newblock Cubic nonlinear squeezing and its decoherence.
\newblock {\em Optics Express}, 30(17):31456--31471, August 2022.

\bibitem{gonzalez-ballestero_suppressing_2023}
C.~{Gonzalez-Ballestero}, J.A. Zieli{\'n}ska, M.~Rossi, A.~Militaru,
  M.~Frimmer, L.~Novotny, P.~Maurer, and O.~{Romero-Isart}.
\newblock Suppressing {{Recoil Heating}} in {{Levitated Optomechanics Using
  Squeezed Light}}.
\newblock {\em PRX Quantum}, 4(3):030331, September 2023.

\bibitem{dago_stabilizing_2024}
Salamb{\^o} Dago, J.~Rieser, M.~A. Ciampini, V.~Mlyn{\'a}{\v r}, A.~Kugi,
  M.~Aspelmeyer, A.~{Deutschmann-Olek}, and N.~Kiesel.
\newblock Stabilizing nanoparticles in the intensity minimum: Feedback
  levitation on an inverted potential.
\newblock {\em Optics Express}, 32(25):45133--45141, December 2024.

\bibitem{mlynar_feedback_2025}
Vojt{\v e}ch Mlyn{\'a}{\v r}, Salamb{\^o} Dago, Jakob Rieser, Mario~A.
  Ciampini, Markus Aspelmeyer, Nikolai Kiesel, Andreas Kugi, and Andreas
  {Deutschmann-Olek}.
\newblock Feedback stabilization of a nanoparticle at the intensity minimum of
  an optical double-well potential, August 2025.

\bibitem{kamba_optical_2022}
Mitsuyoshi Kamba, Ryoga Shimizu, and Kiyotaka Aikawa.
\newblock Optical cold damping of neutral nanoparticles near the ground state
  in an optical lattice.
\newblock {\em Optics Express}, 30(15):26716--26727, July 2022.

\bibitem{bonvin_state_2023}
Eric Bonvin, Louisiane Devaud, Massimiliano Rossi, Andrei Militaru, Lorenzo
  Dania, Dmitry~S. Bykov, Oriol {Romero-Isart}, Tracy~E. Northup, Lukas
  Novotny, and Martin Frimmer.
\newblock State {{Expansion}} of a {{Levitated Nanoparticle}} in a {{Dark
  Harmonic Potential}}, December 2023.

\bibitem{duchan_nanomechanical_2025}
Martin Ducha{\v n}, Martin {\v S}iler, Petr J{\'a}kl, Oto Brzobohat{\'y},
  Andrey Rakhubovsky, Radim Filip, and Pavel Zem{\'a}nek.
\newblock Nanomechanical state amplifier based on optical inverted pendulum.
\newblock {\em Communications Physics}, 8(1):276, July 2025.

\bibitem{rossi_quantum_2024}
Massimiliano Rossi, Andrei Militaru, Nicola~Carlon Zambon, Andreu
  {Riera-Campeny}, Oriol {Romero-Isart}, Martin Frimmer, and Lukas Novotny.
\newblock Quantum {{Delocalization}} of a {{Levitated Nanoparticle}}, August
  2024.

\bibitem{tomassi_accelerated_2025}
Gregoire F.~M. Tomassi, Daniel Veldhuizen, Bruno Melo, Davide Candoli, Andreu
  {Riera-Campeny}, Oriol {Romero-Isart}, Nadine Meyer, and Romain Quidant.
\newblock Accelerated {{State Expansion}} of a {{Nanoparticle}} in a {{Dark
  Inverted Potential}}, March 2025.

\bibitem{mari_gently_2009}
A.~Mari and J.~Eisert.
\newblock Gently {{Modulating Optomechanical Systems}}.
\newblock {\em Physical Review Letters}, 103(21):213603, November 2009.

\bibitem{rakhubovsky_squeezing_2013}
Andrey~A. Rakhubovsky and Sergey~P. Vyatchanin.
\newblock Squeezing of optomechanical modes in detuned {{Fabry-Perot}}
  interferometer.
\newblock {\em Physics Letters A}, 377(18):1317--1322, August 2013.

\bibitem{kustura_mechanical_2022}
Katja Kustura, Carlos {Gonzalez-Ballestero}, Andr{\'e}s de los~R{\'i}os Sommer,
  Nadine Meyer, Romain Quidant, and Oriol {Romero-Isart}.
\newblock Mechanical {{Squeezing}} via {{Unstable Dynamics}} in a
  {{Microcavity}}.
\newblock {\em Physical Review Letters}, 128(14):143601, April 2022.

\bibitem{weinbub_recent_2018}
J.~Weinbub and D.~K. Ferry.
\newblock Recent advances in {{Wigner}} function approaches.
\newblock {\em Applied Physics Reviews}, 5(4):041104, October 2018.

\bibitem{walschaers_nongaussian_2021}
Mattia Walschaers.
\newblock Non-{{Gaussian Quantum States}} and {{Where}} to {{Find Them}}.
\newblock {\em PRX Quantum}, 2(3):030204, September 2021.

\bibitem{lachman_quantum_2022}
Luk{\'a}{\v s} Lachman and Radim Filip.
\newblock Quantum non-{{Gaussianity}} of light and atoms.
\newblock {\em Progress in Quantum Electronics}, 83:100395, May 2022.

\bibitem{rakhubovsky_quantum_2024a}
Andrey~A. Rakhubovsky, Darren~W. Moore, and Radim Filip.
\newblock Quantum non-{{Gaussian}} optomechanics and electromechanics.
\newblock {\em Progress in Quantum Electronics}, 93:100495, January 2024.

\bibitem{hoj_ultracoherent_2021}
Dennis H{\o}j, Fengwen Wang, Wenjun Gao, Ulrich~Busk Hoff, Ole Sigmund, and
  Ulrik~Lund Andersen.
\newblock Ultra-coherent nanomechanical resonators based on inverse design.
\newblock {\em Nature Communications}, 12(1):5766, October 2021.

\bibitem{hoj_ultracoherent_2024}
Dennis H{\o}j, Ulrich~Busk Hoff, and Ulrik~Lund Andersen.
\newblock Ultracoherent {{Nanomechanical Resonators Based}} on {{Density
  Phononic Crystal Engineering}}.
\newblock {\em Physical Review X}, 14(1):011039, March 2024.

\bibitem{bereyhi_perimeter_2022}
Mohammad~J. Bereyhi, Amirali Arabmoheghi, Alberto Beccari, Sergey~A. Fedorov,
  Guanhao Huang, Tobias~J. Kippenberg, and Nils~J. Engelsen.
\newblock Perimeter {{Modes}} of {{Nanomechanical Resonators Exhibit Quality
  Factors Exceeding}} \$\textbraceleft
  10\textbraceright\textasciicircum\textbraceleft 9\textbraceright\$ at {{Room
  Temperature}}.
\newblock {\em Physical Review X}, 12(2):021036, May 2022.

\bibitem{beccari_strained_2022}
A.~Beccari, D.~A. Visani, S.~A. Fedorov, M.~J. Bereyhi, V.~Boureau, N.~J.
  Engelsen, and T.~J. Kippenberg.
\newblock Strained crystalline nanomechanical resonators with quality factors
  above 10 billion.
\newblock {\em Nature Physics}, pages 1--6, February 2022.

\bibitem{gottesman_encoding_2001}
Daniel Gottesman, Alexei Kitaev, and John Preskill.
\newblock Encoding a qubit in an oscillator.
\newblock {\em Physical Review A}, 64(1):012310, June 2001.

\bibitem{johansson_qutip_2012}
J.~R. Johansson, P.~D. Nation, and Franco Nori.
\newblock {{QuTiP}}: {{An}} open-source {{Python}} framework for the dynamics
  of open quantum systems.
\newblock {\em Computer Physics Communications}, 183(8):1760--1772, August
  2012.

\bibitem{johansson_qutip_2013}
J.~R. Johansson, P.~D. Nation, and Franco Nori.
\newblock {{QuTiP}} 2: {{A Python}} framework for the dynamics of open quantum
  systems.
\newblock {\em Computer Physics Communications}, 184(4):1234--1240, April 2013.

\bibitem{lambert_qutip_2024}
Neill Lambert, Eric Gigu{\`e}re, Paul Menczel, Boxi Li, Patrick Hopf, Gerardo
  Su{\'a}rez, Marc Gali, Jake Lishman, Rushiraj Gadhvi, Rochisha Agarwal, Asier
  Galicia, Nathan Shammah, Paul Nation, J.~R. Johansson, Shahnawaz Ahmed, Simon
  Cross, Alexander Pitchford, and Franco Nori.
\newblock {{QuTiP}} 5: {{The Quantum Toolbox}} in {{Python}}, December 2024.

\end{thebibliography}

\end{document}